\documentclass[journal]{IEEEtran}
\usepackage{amssymb,amsmath}
\usepackage{cite}
\usepackage{graphicx}
\usepackage[caption=false]{subfig}
\usepackage[font=footnotesize]{caption}
\usepackage{psfrag}
\usepackage{url}
\usepackage[latin1]{inputenc}
\usepackage[absolute,overlay]{textpos}
\usepackage{tikz}
\usetikzlibrary{arrows,calc,decorations.markings}
\usetikzlibrary{topaths}
\usetikzlibrary{shapes,trees}
\usetikzlibrary{arrows}
\usetikzlibrary{shadows}
\usetikzlibrary{positioning}
\usetikzlibrary{matrix}
\usetikzlibrary{shapes.geometric}
\usetikzlibrary{decorations.pathmorphing}
\usepgflibrary{patterns}
\usetikzlibrary{calc}
\usetikzlibrary{fit}					% fitting shapes to coordinates
\usetikzlibrary{backgrounds}	% drawing the background after the foreground
\DeclareMathOperator*{\argmin}{arg\,min}
\DeclareMathOperator*{\argmax}{arg\,max}
\usepackage{pgf}
\usetikzlibrary{arrows,automata}
\usepackage[latin1]{inputenc}

\usepackage[linesnumbered,ruled,lined]{algorithm2e}
\usepackage{amsthm}

\newtheorem{remark}{Remark}

\usepackage{tabularx}
\usepackage{makecell}
\usepackage{multirow}

\usepackage{amsmath}
\usepackage{amssymb}
\usepackage{bbm}
\usepackage{bm}
\usepackage{comment}
\usepackage[capitalize]{cleveref}

\usepackage{pgfplots}
\usepackage{float}
\usepackage{booktabs}

\pgfplotsset{compat=1.15}
\usetikzlibrary{shapes,positioning,automata, arrows.meta}

\begin{document}

\title{Traffic Prediction and Fast Uplink for Hidden Markov IoT Models}
\author{
	\IEEEauthorblockN{Eslam Eldeeb, Mohammad Shehab, Anders E. Kal\o r, Petar Popovski, and Hirley Alves \\
	}	
	
	\thanks{The work by E. Eldeeb, M. Shehab, and H. Alves has been partially supported by Academy of Finland 6Genesis Flagship (Grant no. 318927), and FIREMAN (Grant no. 326301), and the European Commission through the Horizon Europe project Hexa-X (Grant Agreement no. 101015956). The work by A. E. Kal{\o}r and P. Popovski has been supported by the Danish Council for Independent Research (Grant Nr. 8022-00284B SEMIOTIC) and by the Villum Investigator Grant "WATER" from the Velux Foundation, Denmark.
}
	
	\thanks{E. Eldeeb, M. Shehab, and H. Alves are with Centre for Wireless Communications (CWC), University of Oulu, Finland. Email: firstname.lastname@oulu.fi.}
	
	\thanks{A. E. Kal\o r and P. Popovski are with the Department of Electronic Systems, Aalborg University, Aalborg, Denmark e-mails: {aek, petarp}@es.aau.dk.}
	}
\maketitle

\begin{abstract}
  In this work, we present a novel traffic prediction and fast uplink framework for IoT networks controlled by binary Markovian events. First, we apply the forward algorithm with hidden Markov models (HMM) in order to schedule the available resources to the devices with maximum likelihood activation probabilities via fast uplink grant. In addition, we evaluate the regret metric as the number of wasted transmission slots to evaluate the performance of the prediction. Next, we formulate a fairness optimization problem to minimize the age of information while keeping the regret as minimum as possible. Finally, we propose an iterative algorithm to estimate the model hyperparameters (activation probabilities) in a real-time application and apply an online-learning version of the proposed traffic prediction scheme.
Simulation results show that the proposed algorithms outperform baseline models such as time division multiple access (TDMA) and grant-free (GF) random-access in terms of regret, the efficiency of system usage, and age of information.

%First, we consider a massive set of IoT devices whose activation events are modeled by an On-Off Markov process with known transition probabilities. Next, we exploit the temporal correlation of the traffic events and apply the forward algorithm in the context of hidden Markov models (HMM) in order to predict the activation likelihood of each IoT device. Finally, we apply the fast uplink grant scheme in order to allocate resources to the IoT devices that have the maximal likelihood for transmission. In order to evaluate the performance of the proposed scheme, we define the regret metric as the number of missed resource allocation opportunities. The proposed fast uplink scheme based on traffic prediction outperforms both conventional random access and time division duplex in terms of regret and efficiency of system usage, while it maintains its superiority over random access in terms of average age of information for massive deployments.
\end{abstract}

\section{Introduction}\label{sec:introduction}
%\textcolor{red}{MS: improve organization - highlight the difference between known and unknown parameters}

Recent advances in internet of things (IoT) has led to the deployment of a large number of machine-type communication (MTC) devices to collect real-time information. The number of such IoT-MTC devices is rapidly growing to realize different use cases such as environment monitoring, remote surgery, and autonomous vehicles~\cite{6g}. In 5G, MTC service modes are massive MTC (mMTC) and ultra-reliable low latency communication (URLLC)~\cite{mahmood2019key}. The quality-of-service (QoS) demands vary among the service modes. In addition, many use cases have recently had more strict demands, which need extremely low end-to-end latency in a massive deployment of IoT devices to collect real-time information~\cite{mahmood2019key}.

\begin{figure}[!t] % [!t] or [!b] or [!h] % force fitting, force top, force bottom, force text fitting
		\centering 
		\includegraphics[clip=true,width=0.9\linewidth]{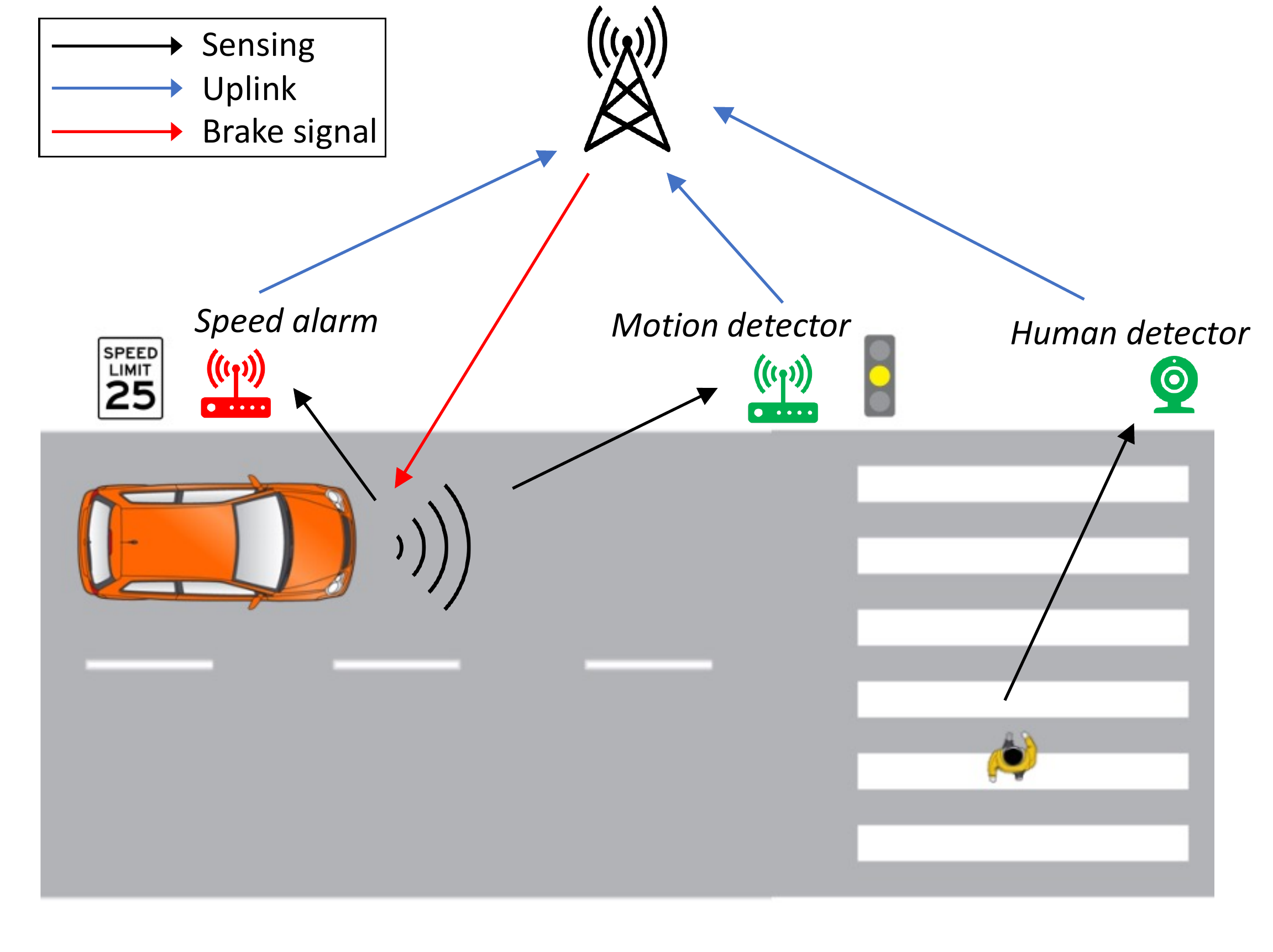}
		\vspace{-0mm}
		\caption{Traffic correlation scenario: A speed alarm would be active only if the motion detector is active but not vice versa. The human detector signal would only be important if the other two sensors are active.}
		\label{FU}
		\vspace{-0mm}
\end{figure}

The behavior of the traffic of MTC devices (MTDs) differs from that of the traditional human-type communication devices (HTDs)~\cite{6629817}. The HTDs traffic tends to be heterogeneous, whereas the traffic of MTDs is homogeneous and highly correlated. To elucidate traffic correlation in MTC, we consider the following road safety example as in Fig.~\ref{FU}: let event 1 and event 2 correspond to a vehicle moving down the street at normal speed, and a vehicle breaking the speed limit, respectively. Meanwhile, sensor 1 and sensor 2 are motion detectors, necessary to control the traffic lights, and speed limit alarm, respectively. In this scenario, event 1 will be detected by sensor 1 only. However, both sensors may likely detect event 2. Hence, we infer that sensor 2 will not likely be active except if sensor 1 is active. Moreover, if sensor 2 is active, sensor 1 will most probably be active but not vice versa. In such a scenario, it is essential to estimate the possible sensor activation pattern and allocate resources at low latency. If a human is crossing the street, a human detector or a road safety alarm could then transmit a signal to the BS. The BS in turn sends a compulsory brake signal to a high-speed vehicle to enforce it to slow down the speed. This all should occur within a window of a few milliseconds to avoid an accident. The importance of an uplink signal from the human detector in this scenario is also dependant on whether the speed alarm is active or not.

Another example to illustrate traffic correlation, let Markovian event 1 and Markovian event 2 correspond to the existence of fire or no fire, and someone who smokes a cigarette or no smoke, respectively. Meanwhile, sensor 1 and sensor 2 are heat and smoke detectors, respectively. In case of fire, both sensors will detect the event. However, in case of smoking a cigarette, event 2 will only be detected by sensor 2. Hence, we infer that if sensor 1 is active, sensor 2 will be active with high probability but not vice versa.

%let event 1 and event 2 correspond to the existence of fire, and someone who smokes a cigarette, respectively. Meanwhile, sensor 1 and sensor 2 are heat and smoke detectors, respectively. In case of fire, both sensors will detect the event. However, in case of smoking a cigarette, event 2 will only be detected by sensor 2. Hence, we infer that if sensor 1 is active, sensor 2 will be active with high probability but not vice versa. \AK{This example is not ideal, because there is no need that the sensor 2 is ever active. Also, we mention latency but heat/smoke is not very latency critical. Can we maybe think of some robotics scenario instead?}

One important metric to measure the freshness of received data from an IoT device is the age-of-information (AoI). AoI was first introduced in~\cite{kaul2011minimizing}. It defines the freshness of information (time elapsed since data at a source has been collected and transmitted to a destination). Therefore, minimizing the AoI in IoT networks has become essential when designing scheduling algorithms~\cite{bedewy}. 

A key element in communication systems is the design of access protocols, which allow the devices to transmit their data in an organized manner. In what follows, we discuss the shortcomings of existing massive access protocols. In conventional LTE systems, the devices communicate with the base station (BS) using the random access (RA) procedures~\cite{8705373}, e.g., each device goes through a 4-handshake procedure initiated by the transmission of a random preamble followed by a random-access response from the BS side. Afterward, the device requests a connection and the BS responds with a contention resolution message. However, this procedure suffers from high signaling overhead and end-to-end latency, which fails to serve strict low latency demands and results in a relatively high AoI. %\AK{Elaborate RA procedure, otherwise it is difficult to see why it has high overhead and latency}.
Furthermore, due to the limited number of preambles, it is susceptible to a high number of collisions in situations where a large number of devices sporadically try to access the network at the same time, such as in an alarm scenario~\cite{RA}.

Meanwhile, alternative solutions have been proposed to solve the problems of collisions and signaling overhead in IoT networks, from legacy time division multiple access (TDMA) to grant-free (GF) schemes, and access class barring (ACB). In TDMA schemes, the resources are distributed equally among the devices without considering any scheduling algorithms. Although TDMA is straightforward and efficient in periodic transmission scenarios, it does not perform well when the traffic is sporadic and event-driven. Therefore, GF access has been proposed as an efficient procedure to reduce the signaling overhead by skipping the preamble request and reply that constitute the first 2-steps of the 4-handshake procedure~\cite{8877253}. %Although GF solutions reduce the signaling overhead to half of the grant-based RA, it fails when the number of potentially active devices exceed the available resources as it suffers from large number of collisions, which cause high AoI experienced by the devices.
Although GF solutions reduce the signaling overhead to half of the grant-based RA, it fails when the number of potentially active devices exceeds the available resources. In addition, it suffers from a large number of collisions, which cause high AoI experienced by the devices.
Among the alternative solutions, promising results have been obtained for ACB~\cite{ACB_INTRO}. The device generates a random number between 0 and 1 and compares it with the ACB factor broadcasted by the BS. The device can only access the BS if the generated number is less than the ACB factor. Although the literature has a vast amount of works extending the basic idea of the ACB, such as extended ACB~\cite{7029666}, cooperative ACB~\cite{6093905}, and dynamic ACB~\cite{6855701}, ACB still fails to satisfy strict latency requirements~\cite{ACB}.
To this end, the need of extreme low latency in IoT urges the design of novel access schemes to overcome the flaws associated with the old ones. Learning-based schemes were discussed in many surveys as the potential solution to the existing problems of the proposed approaches in the literature to overcome the RA limitations~\cite{9355403}. Moreover, many emerging IoT applications can exploit activation correlation and traffic prediction to enable pre-emptive resource allocation and achieve ultra-reliable and low latency communications. The traffic correlation behavior of MTDs enables traffic prediction and forecasting algorithms to anticipate the set of active and silent MTDs. 
In this context, Fast Uplink (FU) grant was introduced in~\cite{FU} to allow for resource allocation based on traffic prediction schemes.

\subsection{Fast Uplink Grant}

To elaborate more on FU grant, we consider $K$ IoT devices and $L$ available transmission slots, where $K\gg L$. Each device is stimulated to generate data packets at different time slots controlled by different processes at the application layer, e.g., triggered external events. Whenever a device generates a packet, it will need a transmission slot to transmit it to the BS. In the FU scheme, the BS allocates the available transmission slots to the set of IoT devices that it believes will transmit in the current time slot. The designed resource allocation scheme should exploit the correlation of traffic pattern based on the temporal and event dimensions.

%An FU grant scenario is illustrated in Fig. \ref{FU} with $K=5$ devices and $L=2$ slots at a certain time instant, where only 2 out of 5 devices are active. 
%
%The FU grant scenario relies mainly on traffic prediction, where the BS has to predict efficiently the probability of each device to be active or silent and grant the available resources to those most likely to be active, with some fairness guarantees. Some of the potential advantages of applying FU grant are:
The FU scenario relies mainly on traffic prediction. The BS has to efficiently predict the probability of each device to be active or silent and grant the available resources to those most likely to be active, with some fairness guarantees. Some of the potential advantages of applying FU are:
\begin{itemize}
\item Absence of scheduling requests and collisions leading to a reduction in the energy consumption of IoT devices and uplink latency;
\item Clearance of signaling overhead between the devices since learning occurs only at the side of the BS;
\item It allows for the potential use of the uplink grant signal to partially or fully estimate the channel condition at the IoT devices side before actual uplink process (CSIT)\footnote{%Note that the channel estimation is a proposed advantage of applying the FU grant scheme that could be investigated for possible future implementation. This could be performed for example, via pilot symbols transmitted within the uplink grant signal.
Notice that channel estimation is a proposed advantage when applying the FU scheme, e.g., via pilot symbols transmitted within the uplink grant signal. However, we leave this work for future implementation.
}.
\end{itemize}

%\AK{Some transition is needed here. Also, we need to provide a short introduction of our idea somewhere before the related work so that we can connect it to our contribution.}

%\begin{figure}[!t] % [!t] or [!b] or [!h] % force fitting, force top, force bottom, force text fitting
%		\centering 
		%\includegraphics[clip=true,width=1\linewidth]{Figures/FUG_illustr.pdf}
		%\vspace{-0mm}
		%\caption{Fast Uplink scenario. The BS grants access to the predicted active devices only.}
		%\label{FU}
		%\vspace{-0mm}
	%\end{figure}

\subsection{Contributions}
In this work, we build upon~\cite{9217258}, where we define the main system model that consists of a set of binary discrete events that affects the activation patterns of massive IoT devices. The binary events are modeled as Markovian sources.
%We introduce an FU grant algorithm that exploits the traffic correlation to predict the traffic pattern efficiently of the IoT devices using HMM and the forward algorithm. 
We introduce an FU algorithm that exploits the traffic correlation to efficiently predict the IoT devices' traffic pattern using hidden Markov model (HMM) and the forward algorithm. 
The forward algorithm is a learning algorithm that fits the proposed HMM. The results show that the FU algorithm outperforms the conventional RA and TDMA schemes in terms of the accuracy and efficiency of resource allocation.

Another novel contribution is that we post-process the prediction of the forward algorithm to lower the average experienced for AoI all devices at each time step while maintaining the prediction accuracy as high as possible. We optimize an age parameter to increase the resulting allocation index of the high-age devices and guarantee a higher degree of scheduling fairness. In addition, we formulate a baseline model based on the forward algorithm that forms a distribution of the activation probability using extremely low computation resources. %Furthermore, we estimate the model hyperparameters to exploit the formulated FU grant algorithm in real-time applications, where there are no prior knowledge of the model hyperparameters. 
Furthermore, we estimate the model hyperparameters to exploit the formulated FU algorithm in real-time applications without prior knowledge of the model hyperparameters. 
We then propose an online-learning version of the FU algorithm, where the BS exploits only the set of observations at each instant to allocate the resources to the devices using the learnt hyperparameters. The simulation results illustrate that applying the online-learning algorithm at each instant still captures the age and the accuracy of the actual genie-aided model and outperforms the traditional resource allocation schemes and the HMM baseline scheme.

The contributions of this work are summarized as follows:
\begin{itemize}
    \item We formulate the device activation probabilities for the described HMM system model.
    
    \item We apply the forward algorithm to predict the active devices and perform preemptive FU grant with low complexity.
    
    \item We optimize an age parameter to compensate the AoI of the devices that have experienced high AoI while preserving the accuracy of the efficient forward algorithm. 
    
    \item For the case of unknown hyperparameters of the model, we apply an expectation-maximization algorithm to estimate the event transition probabilities and the device activation probabilities based only on the observations. Then, we apply the estimation procedure to present an offline-learning version of the FU algorithm.
    
    \item Finally, we rely on both the AoI compensation and the learned parameters to formulate an online-learning scheme that allows the BS to perform the FU algorithm in real-time applications, without prior availability of large activation data sets.

    \item The proposed online and offline schemes clearly outperform conventional GF and TDMA in terms of resource allocation efficiency while guaranteeing a favourable amount of fairness via age compensation. 
\end{itemize}

\subsection{Outline}
The rest of the paper is organized as follows: Section~\ref{sec:related_work} discusses the related literature. Section~\ref {sec:sysmodel} depicts the system model for the IoT device. It also explains performance metrics that are used to evaluate the performance of the proposed FU schemes. Next, Section~\ref{sec:sysanalysis} applies the forward algorithm to predict the traffic pattern of IoT devices. After that, Section~\ref{ONLINE_LEARNING_SUBSECTION} discusses the online-learning version of the FU algorithm. Section~\ref{results} depicts and discusses different results for the performance evaluation. Finally, Section~\ref{conclusions} concludes the paper and discusses future research directions. 

\textbf{Notation:} Boldface lowercase letters denote vectors. $Pr$ denotes the probability equation. In addition, $[x]^+$ refers to $\max(0,x)$, $\argmax$ is the maximization notation, and $\argmin$ is the minimization notation. $\Bar{x}$ is the mean of $x$ and $C(a,b)$ is the cost function, where $a$ and $b$ are the parameters to be optimized. To make the paper more tractable, we summarize the key abbreviations and symbols that will appear throughout the paper in Table~\ref{TABLE_I}.

\begin{table}[!ht]
	\centering
	\caption{Important abbreviations and symbols.}
	\label{TABLE_I}
	\renewcommand{\arraystretch}{1.3}
	\begin{tabular}{p{1.5cm} p{5cm}}
		\hline
        
        ACB & access class barring \\
        AoI & age-of-information \\
        CMAP & coupled Markovian arrival process \\
        CMMPP & coupled Markov modulated Poisson process\\ 
        DRL & deep reinforcement learning \\
        FU & fast uplink \\
        GF & grant-free \\
        HMM & hidden Markov model  \\
        LSTM & long short-term memory \\
        MTD & machine-type communication device \\
		NOMA & non-orthogonal multiple access \\
		PDF & probability density distribution \\
		RA & random access \\
		RNN & recurrent neural network \\
		SVM & support vector machine \\
		TDMA & time division multiple access \\
		\vspace{8mm} \\

		$K$ & number of IoT devices \\
		$L$ & number of frequency resources \\
		$N$ & number of Markovian events \\
		$Z$ & number of Baum-Welsh iterations \\

	    \vspace{8mm} \\
	    
	    $\epsilon_0$, $\epsilon_1$ & temporal transition probabilities \\
	    $q_{nk}$ & activation probabilities \\
	    $\omega_t$ & wrong allocations \\
	    $\mu_t$ & missed allocations \\
	    $\beta$ & age parameter \\
	    $I_{t+1}^{(k)}$ & scheduling priority index \\
	    $\Bar{R}$ & average regret \\
	    $\Bar{\Delta}$ & average age \\
	   	\vspace{8mm} \\
	 \hline
	\end{tabular} 
\end{table}

\section{State of the art} \label{sec:related_work}
Many learning-based schemes have been proposed in the literature for resource allocation in IoT networks. In this section, we present a brief literature review of the existing schemes and discuss their limitations. To begin with, in~\cite{CM1,CM2}, the authors studied the activation of devices following coupled Markov modulated Poisson process (CMMPP) and coupled Markovian arrival process (CMAP) traffic models, respectively. However, they did not offer resource allocation schemes based on these traffic models. In~\cite{7061424}, the authors used an HMM model to build a decision fusion algorithm that investigates the correlation time between binary sources in a wireless sensor network (WSN). In the same context, the work in~\cite{Anders_RA} exploited the correlated activity of devices to develop heuristic protocols for GF RA. Sinusoidal spreading sequences were proposed in~\cite{hasan2020fast} to enable FU grant based on free non-orthogonal multiple access (NOMA), whereas authors in~\cite{8417634} discussed hybrid resource allocation schemes to overcome the large signaling overhead and collision problems resulting from message replications in GF transmission. Moreover, in~\cite{Samad_MAB}, Samad et al. introduced a multi-armed bandit algortihm to perform FU grant in IoT networks. However, this work also came short from exploiting the traffic correlation on the event-temporal basis.

The authors in~\cite{9504554} present an FU grant algorithm based on support vector machines (SVM) and long short-term memory (LSTM). However, the addressed algorithm needs efficient hardware at the BS to carry out complex neural networks computations. Authors in~\cite{Federated_Learning} presented an FU grant-based federated learning approach, where the BS relies on the traffic estimation at the side of the devices. Although performing the estimation at the side of the devices side reduces the complexity at the BS side, which is responsible only to perform allocation, it requires the low power end devices to perform complex computations. In addition, authors in~\cite{7511617} formulate a reinforcement learning algorithm for resource allocation in device-to-device (D2D) communications, whereas authors in~\cite{kalor2021prediction} propose a recurrent neural network (RNN) model based on meta-learning to predict the millimeter wave (mmWave) link blockages. Mohammadi et al.~\cite{8403658} presented a multi-agent deep reinforcement learning (DRL) solution for resource allocation, authors in~\cite{8972358} proposed a clustering-based solution to perform resource scheduling depending on each cluster priority and demands, and the work in~\cite{9628555} presented a survey of recent artificial intelligent (AI)-based frameworks for resource allocation in diverse use cases. Table~\ref{TABLE_II} summarizes the existing reviewed literature.

The majority of the referred literature relies on the use of machine learning and reinforcement learning schemes, which need to perform complex computations either at the BS side or the IoT devices side. This requires powerful hardware and a long training duration that reflects some challenges on the usage of machine learning in communication systems~\cite{9061001}. %\AK{This is not a very strong argument. We need to emphasize either that the training is difficult (but do we improve wrt. that?), or that our method provides better insight into the performance, or something else.} 
In addition, IoT networks are often driven by interactive applications, where observations are provided based on human/machine interaction over time, which means that adding a set of new observations to the collected observations for a period of time changes the model and the learning problem. Therefore, online learning becomes necessary~\cite{ONLINE_LEARNING_CITE}. Hence, generalized complex machine learning schemes might not be able to train real-time IoT networks as they require extremely powerful hardware to perform their learning algorithms online and simpler, specially tailored, learning schemes are required for online learning scheduling algorithms~\cite{8449070}. In this work, we present a stochastic-based solution, which fits well with the proposed HMM model. Moreover, it is very efficient in terms of prediction accuracy and simpler than the existing machine learning solutions in the literature.

\begin{table}[!t]
\centering
    \caption{Sumary of the literature review.}
	\label{TABLE_II}
\begin{tabular}{|l|l|l|}
\hline
Main scope & Literature & Sub-topics \\\hline
\multirow{3}{*}{Traffic Models} & M. Laner et al.~\cite{CM1} & CMMPP
\\ \cline{2-3}
 & E. Grigoreva et al.~\cite{CM2} & CMAP
\\ \cline{2-3}
 & P. Rossi et al.~\cite{7061424} & HMM in WSN 
\\ \Xhline{2\arrayrulewidth}

\multirow{3}{*}{RA-based} & A. E. {Kal{\o}r} et al.~\cite{Anders_RA} & GF-RA
\\ \cline{2-3}                               
 & S. M. Hasan et al.~\cite{hasan2020fast} & NOMA FU grant\\ \cline{2-3}                    
 
 & Z. Zhou et al.~\cite{8417634} & Hybrid resource allocation
 \\ \Xhline{2\arrayrulewidth}

\multirow{3}{*}{FU grant} & S. Ali et al.~\cite{Samad_MAB} & Multi-armed bandit
\\ \cline{2-3}                               
 & E. Eldeeb et al.~\cite{9504554} & SVM and LSTM\\ \cline{2-3}                    
 
 & O. Habachi et al.~\cite{Federated_Learning} & Federated learning  \\ \Xhline{2\arrayrulewidth}

\multirow{5}{*}{Deep learning} & I. AlQerm et al.~\cite{7511617} & Reinforcement learning
\\ \cline{2-3}                               
 & A. E. {Kal{\o}r} et al.~\cite{kalor2021prediction} & Meta-learning and RNN\\ \cline{2-3}                    
 
 & F. Mohammadi et al.~\cite{8403658} & Multi-agent DRL  \\ \cline{2-3}    

 & X. Liu et al.~\cite{8972358} & Clustering  \\
 \cline{2-3}
 
 & D. Hejji et al.~\cite{9628555} & AI survey \\
\hline                              
\end{tabular} \vspace{-0mm}
\end{table}

\section{System model and Problem Formulation}\label{sec:sysmodel}
\begin{figure}[!t] % [!t] or [!b] or [!h] % force fitting, force top, force bottom, force text fitting
\centering 
\includegraphics[trim=3cm 13.5cm 3cm 1cm, clip=true,width=1\linewidth]{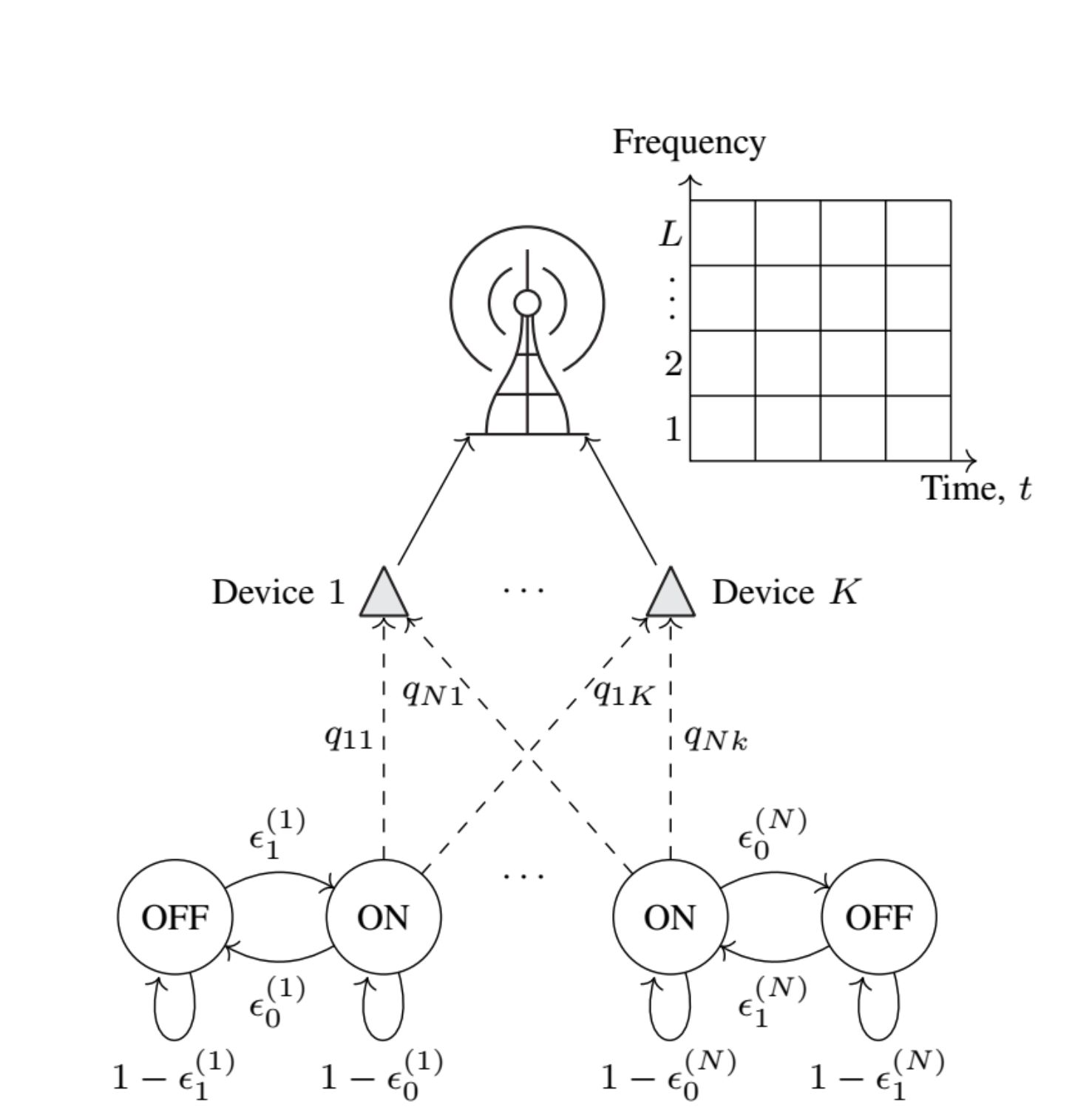}
%\vspace{-2mm}
\caption{The considered activation model, in which $N$ On-Off Markovian processes control the activation of $K$ devices. If process $n$ is in the On-state it activates device $k$ with probability $q_{nk}$.}
\label{fig:scenario}
%\vspace{-2mm}
\end{figure}

Consider an IoT network, such as NB-IoT, with $K$ IoT devices relay their information to a single BS as depicted in \cref{fig:scenario}. As in conventional LTE FU, the transmission resources are link is divided into time slots, and in every time slot the BS can schedule up to $L$ devices for transmission in $L$ frequency slots. The scheduled devices are assigned to transmission slots and they transmit only if they are active (i.e. if they have data to transmit). 
%, and in that case a transmission may fail with probability $p_{\mathrm{err}}$}.
If a device is scheduled for transmission while inactive, the uplink resource is wasted.

We denote the activation of device $k$ in discrete time slots $t=1,2, \dots$ by the random variable $A_t^{(k)}$. $A_t^{(k)}=1$ if the device is active, otherwise $A_t^{(k)}=0$. The activation of IoT devices at time $t$ is indicated by the vector $\textbf{A}_t = \left\lbrace A_t^{(1)}, ..., A_t^{(K)} \right\rbrace$.

\subsection{State Transition Probabilities}
The activation of the devices is controlled by $N$ independent two-state Markov processes. The Markovian processes swing between On and Off states, where at time $t$, the state $\mathcal{S}_t^{(n)} \in \ \left\lbrace 1,0\right\rbrace $, is governed by temporal transition probabilities $\epsilon_1^{(n)}$, $\epsilon_0^{(n)}$ as shown in Fig.~\ref{fig:scenario}, where
\begin{align}
    \Pr\left( \mathcal{S}_{t+1}^{(n)}=0 \middle| \mathcal{S}_t^{(n)}=1\right) &=\epsilon_0^{(n)}, \\
    \Pr\left( \mathcal{S}_{t+1}^{(n)}=1 \middle| \mathcal{S}_t^{(n)}=0\right) &=\epsilon_1^{(n)}, \\
    \Pr\left( \mathcal{S}_{t+1}^{(n)}=0 \middle| \mathcal{S}_t^{(n)}=0\right) &=1-\epsilon_1^{(n)},  \\
    \Pr\left( \mathcal{S}_{t+1}^{(n)}=1 \middle| \mathcal{S}_t^{(n)}=1\right) &=1-\epsilon_0^{(n)}.
\end{align}
%%\vspace{-3mm}
%\begin{equation}
%    \Pr\left( \mathcal{S}_{t+1}^{(n)}=1 \middle| \mathcal{S}_t^{(n)}=0\right) =\epsilon_1^{(n)}, 
%\end{equation}
%\begin{equation}
%    \Pr\left( \mathcal{S}_{t+1}^{(n)}=0 \middle| \mathcal{S}_t^{(n)}=0\right) =1-\epsilon_1^{(n)}, 
%\end{equation}
%\begin{equation}
%    \Pr\left( \mathcal{S}_{t+1}^{(n)}=1 \middle| \mathcal{S}_t^{(n)}=1\right) =1-\epsilon_0^{(n)}. 
%\end{equation} 
%%\vspace{-3mm}
%
To this end, we define the state vector at time $t$ as $\mathcal{\textbf{S}}_t=\left\lbrace \mathcal{S}_t^{(1)},...,\mathcal{S}_t^{(N)}\right\rbrace $.
The Markov processes that are in the On state, i.e. $\mathcal{S}_t^{(n)}=1$, may activate specific IoT devices, where the probability that Markov process $n$ activates device $k$ is given by $q_{nk}$.

%%\vspace{2mm}

\subsection{Device Activation Probabilities}
A certain device becomes active if one or more of the Markovian states activates it. Thus, the probability that device $k$ is active at time $t$ is 
\begin{align}\label{activation_Prop}
   \Pr\left( A_{t}^{(k)}=1 \middle| \mathcal{\textbf{S}}_t\right)&=1-\bigcap_{n=1}^N \Pr\left( A_{t}^{(k)}=0 \middle | \mathcal{S}_t^{(n)}  \right) \\
   &=1-\prod_{n=1}^N (1-q_{nk})^{\mathcal{S}_t^{(n)} },
\end{align}
where the activation is considered to be conditionally independent given the state vector $\mathbf{S}_t$.

Furthermore, the probability that IoT device $k$ will be active at the future time instant $t+1$ given the state vector at time $t$ can be written as
\begin{align} \label{activation_t_1}
   \Pr\left( A_{t+1}^{(k)}=1 \middle| \mathcal{\textbf{S}}_t\right)&=1-\bigcap_{n=1}^N \Pr\left( A_{t+1}^{(k)}=0 \middle | \mathcal{S}_t^{(n)}  \right) \\
   &=1-\prod_{n=1}^N h(n),
\end{align}
where
\begin{equation}
   h(n)=
      \begin{cases}
        1-\epsilon_1^{(n)}+\epsilon_1^{(n)}(1-q_{nk}), & \quad \mathcal{S}_t^{(n)}=0\\
       \epsilon_0^{(n)}+(1-\epsilon_0^{(n)})(1-q_{nk}), & \quad  \mathcal{S}_t^{(n)}=1.
     \end{cases}
   \end{equation}

\subsection{Performance Evaluation Metrics}
Next, we define key performance metrics that are essential to evaluate the proposed FU scheme with traffic prediction and compare it to existing allocation schemes.

\subsubsection{Regret}
The \emph{regret} is one of the key metrics used to evaluate the performance of scheduling algorithms using learning schemes~\cite{FU}.
%
%We define one unit of regret as wasting a resource on an inactive device, while there is one active device that did not receive a resource. 
We define one unit of regret as wasting a resource on an inactive device while one active device did not receive a resource.
Therefore, regret is %defined as 
the accumulated regret units at each time slot that resulted from the prediction and scheduling of active devices. Consider the uplink grant vector $\mathbf{U}_t=\left\lbrace u_t^{(1)}, \dots, u_t^{(K)}\right \rbrace$, where $u_t^{(k)}=1$ if a slot is allocated to device $k$ at time $t$ and $u_t^{(k)}=0$ if device $k$ does not receive a transmission slot. The number of wrong allocations at time instant $t$ can be calculated as the difference between the uplink grant vector $u_t^{(k)}$ at time instant $t$ and the activation vector $A_t^{(k)}$ at time instant $t$ as follows
%
%Here, regret can be defined as the missed opportunity that occurs when a slot is allocated to an inactive device, while there is another device that has data to transmit but receives no grant. We consider the problem of predicting and subsequently scheduling the active devices in each time slot. To this end, regret is cumulative number of wasted resources that could have been allocated to active unserved devices. 
%Hence, the regret can be considered as an indicator of the efficiency of the system usage. In order to capture the scheduling decisions at time $t$, consider the uplink grant vector $\mathbf{G}_t=\left\lbrace g_t^{(1)}, \dots, g_t^{(K)}\right \rbrace$, where $g_t^{(k)}$ is equal to 1 if a slot is allocated to device $k$ and 0 if no slot is allocated to device $k$. In this case, the number of wrong allocations is defined as the number of slots that are allocated to devices, that have no data to transmit which is
%
\begin{equation}
    \omega_t=\sum_{k=1}^K \left[ u_t^{(k)}-A_t^{(k)}\right]^+,
\end{equation}
where $[x]^+=\max(0,x)$.
In addition, the number of missed allocations can be computed as the difference between the activation vector $A_t^{(k)}$ at time instant $t$ and the uplink grant vector $u_t^{(k)}$ at time instant $t$ as follows:
\begin{equation}
    \mu_t=\sum_{k=1}^K \left[ A_t^{(k)}-u_t^{(k)}\right]^+.
\end{equation}
Hence, the regret function at time $t$ is defined as %\vspace{0.5mm}
\begin{equation}
    R(t) = \min \left \lbrace \omega_t,\mu_t \right \rbrace.
\end{equation}
Then, minimizing the long-term $R(t)$ is an important target, when designing an FU grant scheme.

The meaning of the regret function can be understood by considering the following three cases. First, if $M>L$ devices are active and all the $L$ uplink grants are given to a subset of the active devices, then $\omega_t=0$ and $\mu_t=0$. This results in a regret of $R(t)=0$, reflecting that the number of unserved devices is minimized. If no devices are active, and the $L$ grants are given to inactive devices, $\omega_t=L$ and $\mu_t=0$. This also results in $R(t)=0$, again reflecting a minimum number of unserved devices. Finally, if $M\le 2L$ devices and scheduler assigns grants to $M/2$ of the active devices and $L-M/2$ inactive devices, then $\omega_t=L-M/2$ and $\mu_t=M/2$. The regret is then $R(t)=\min(L-M/2,M/2)$, which renders the number of unserved devices that could have been served if the allocation process was more accurate.
%%\vspace{1mm}

\subsubsection{System usage} %%\vspace{1mm}
We propose the system usage metric which would help with evaluation the efficiency of the proposed FU grant allocation scheme. The average system usage $\eta_t$ at time $t$ is defined as the ratio between the number of transmission slots that are successfully used by an IoT device to the total number of available slots $L$ averaged over time. That is 
\begin{equation}
    \eta_t=\frac{1}{tL}\sum_{\tau=0}^t L-\omega_\tau.
    \label{USAGE_EQ}
\end{equation}
The average system usage marks the percentage of transmission slots that are successfully used for uplink by the IoT devices.
%\vspace{1mm}

\subsubsection{Age of Information} %\vspace{1mm}
To measure the freshness of data and the degree of fairness in scheduling the devices, we define the discrete AoI~\cite{AoI,bedewy} of device $k$ as the time passed since the device transmitted a packet. That is the last time instant in which device $k$ was active and received a transmission grant and %%\vspace{-2mm}
\begin{equation}
\Delta^{(k)}=t-t_k,
\end{equation} %\vspace{-2mm}
where $t_k<t$ is the last time slot before $t$, when $A_{t_k}^{(k)}=u_{t_k}^{(k)}=1$ and the AoI should be a non-negative integer. The average age per device at a certain time is defined as %
\begin{equation}
    \Bar{\Delta}=\frac{1}{K}\sum_{k=1}^K \Delta^{(k)}.
\end{equation}
Meanwhile, the peak age per device can be noted as $\max_k \lbrace \Delta^{(k)} \rbrace$.

AoI is important in the proposed scenario since it provides a measure for the freshness of the data received from each IoT device. This means that if a device is rarely scheduled for transmission, the information stored at the BS from this device will be outdated as the device's age becomes too high.% This could be considered as a measure of fairness where devices that are fairly scheduled from time to time would have a relatively low average age.
Hence, it is also considered as a measure of fairness, where higher average ages mean that some devices are rarely scheduled and low average age means that devices are fairly scheduled.

\begin{remark}
We assume that the BS has pre-knowledge of the environment, and hence knows the state transition probabilities $\epsilon^{(n)}$ and the device activation probabilities $q_{nk}$. %The BS aims to jointly minimize the regret and the AoI and maximize the system usage resulting from scheduling the set of available transmission resources to the devices.
Therefore, the BS aims to jointly minimize the regret and the AoI and maximize the system usage by scheduling the available transmission resources to the devices.
In addition, we investigate the same objective while assuming that the state transition probabilities and the device activation probabilities are not fully known by the BS. Hence, the BS needs to estimate the model hyperparameters via estimation algorithms.
\end{remark}

\section{The Proposed Fast Uplink Algorithm}\label{sec:sysanalysis} %\vspace{1mm}
%In this section, we analyze the device temporal activation probabilities and exploit them to develop the traffic prediction based FU scheme. 
This section analyzes the device's temporal activation probabilities and exploits them to develop the traffic prediction-based FU scheme.
The BS uses the set of past observations of each device to predict the hidden states for each event. Afterward, it uses the set of predicted hidden states to generate an estimate for the future observations for each device. 

\subsection{Traffic Prediction}

The BS does not know the states of the Markov processes and hence, continuously needs to estimate them based on the observations. Notice that the activation process of the IoT devices can be described by an $N$-HMM as typically detailed in~\cite{HMM}. Concretely, the forward algorithm can be applied by the BS to learn the probability of events being in a certain state given the history of IoT devices activation observations done by the BS~\cite{9409161}. The BS can exploit the learned state distribution to estimate future device activation probabilities and patterns.

To obtain a clear understanding of the forward algorithm, consider the joint probability $p(\mathbf{S}_t, \mathbf{A}_t)$. The forward algorithm is able to efficiently compute this joint probability in a recursive way as in~\cite{HMM2}. Herein, the forward algorithm is described as follows
\begin{equation} \label{alpha_equation}
p(\mathcal{\textbf{S}}_t,\textbf{A}_{1:t}) = p\left( \textbf{A}_{t} \middle| \mathcal{\textbf{S}}_t \right) \sum_{\textbf{S}_{t-1}} p\left( \textbf{S}_t \middle|  \textbf{S}_{t-1} \right) p(\mathbf{S}_{t-1}, \mathbf{A}_{1:t-1}).
\end{equation}
Then the most likely hidden state for the events can be learned using
\begin{equation}\label{eq:most_likely_S}
\textbf{S}_t^*=\argmax_{\textbf{S}_t} ~ p(\mathcal{\textbf{S}}_t,\textbf{A}_{1:t}).
\end{equation}
The estimated hidden states at time instant $t$ are used to predict the activation probabilities of each device at time instant $t+1$ using~\eqref{activation_t_1}. The predicted device activation probabilities can be formulated as
\begin{align} \label{Activation_Prediction}
   \Pr\left( A_{t+1}^{*(k)}=1 \middle| \mathcal{\textbf{S}}_t^*\right)&=1-\bigcap_{n=1}^N \Pr\left( A_{t+1}^{(k)}=0 \middle | \mathcal{S}_t^{*(n)}  \right).
\end{align}

Alternatively, the BS can use the forward algorithm results directly to predict the maximum likelihood of the pattern of the devices in the next time instant
\begin{align}\label{eq:most_likely_A}
\textbf{A}_{t+1}^*&=\argmax_{\textbf{A}_{t+1}}   \sum_{\mathbf{S}_t}\Pr\left( \mathbf{A}_{t+1} \middle| \textbf{S}_t \right) p(\mathcal{\textbf{S}}_t,\textbf{A}_{1:t})\\
&=\argmax_{\textbf{A}_{t+1}}   \sum_{\mathbf{S}_t}p(\mathcal{\textbf{S}}_t,\textbf{A}_{1:t})\prod_{k=1}^K \Pr\left( A_{t+1}^{(k)}=b_k \middle| \textbf{S}_t \right),
\end{align}
where $\textbf{A}_{t+1}^*$ is the maximum likelihood estimate of the set of active IoT devices at time $t+1$, and $b_k \in \ \left\lbrace 1,0\right\rbrace$.% Equation \eqref{eq:most_likely_A} can be iteratively used to predict the state of events at $t+2$ using \eqref{eq:most_likely_S} after partial correction of the activation pattern by setting the activation indicator $A_{t+1}^{(k)^*}$ to zero (limited information) for devices that were granted transmission but did not transmit.

%\AK{I don't understand the previous paragraph. I suggest instead to remove $=b_k$ from the equation and write: Note that \eqref{eq:most_likely_A} and \eqref{eq:most_likely_S} can be iteratively applied to predict the state at any future time $t'\ge t+1$.}

%The target of this algorithm \cite{HMM2} is to predict the IoT devices activation pattern $\textbf{A}_{t+1}$ at time slot $t+1$ based on the activation pattern observations $\textbf{A}_{t}$ at time $t$. The idea is quite close to the Viterbi algorithm where the algorithm determines the most likely activation sequence of the Markovian processes that leads to the current observations. Then the algorithm attempts at predicting the IoT devices activation pattern at the next time slot by maximizing the conditional probabilities given in \eqref{device_prediction} as follows

%\AK{The following paragraph should be moved out of prediction sub-section. Maybe to regret, maybe to its own sub-section after regret.}
Note that~\eqref{eq:most_likely_A} evaluates the probability of a full pattern. Hence, it gives the most likely activation pattern and does not consider the activation probability of each device separately. Meanwhile, when performing uplink grant allocation, the BS should select the $L$ devices which are most likely to be \emph{jointly} active. In order to determine these devices, we assume that the system is in the most likely state, found from~\eqref{eq:most_likely_S}, and exploit this assumption to compute the transition probability of the events as follows
\begin{equation}
    P_{\text{On}}^{(n)} = \Pr \left (\mathcal{S}_{t+1}^{*(n)} = 1 \middle | \mathcal{S}_{t}^{* (n)} = 1 \right ),
\end{equation}
which will be used to determine the activation likelihood of each device as
\begin{align}
    P_{\text{device}}^{(k)} = P_{\text{On}}^{(1)} \cdot q_{1,k} & \bigcup\\
    P_{\text{On}}^{(2)} \cdot q_{2,k} & \bigcup\\
    \vdots \: \: \: \: \: \: \: & \bigcup \\
    P_{\text{On}}^{(n)} \cdot q_{n,k} & .
\end{align}
%or alternatively, we can use the same notation as in \eqref{activation_t_1} and apply the most likely hidden states to predict the probability of a device to be active at the next time instant
%\begin{equation}
%    \Pr\left( A_{t+1}^{(k)}=1 \middle| \mathcal{\textbf{S}}_t^*\right)=1-\bigcap_{n=1}^N \Pr\left( A_{t+1}^{(k)}=0 \middle | \mathcal{S}_t^*{(n)}  \right).
%\end{equation}
Finally, the devices are sorted by their activation probability, and the $L$ devices most likely to be active are scheduled in the next slot.

%\begin{equation}\label{e_activation prediction}
%\hat{\Pr}\left( A_{t+1}^{(k)}=1 \middle| \mathbf{S}_t \right)= 1 - \prod_{n=0}^{N-1} 1 - (q_{nk} \cdot \mathcal{S}^{(n)}_t)
%\end{equation}
%For the vector $\mathbf{\hat{P}}_{t+1}$ whose $k^{th}$ element is equal to the estimated activation probability $\hat{\Pr}\left( A_{t+1}^{(k)}=1\right)$ of device $k$, the uplink grant is given to the $L$ highest elements of this vector. That is, we schedule the $L$ devices with the highest estimated activation probability at a certain time slot $t+1$.
 
%%\vspace{2mm}
%\section{Extensions}\label{extensions}

\subsection{Baseline Model}\label{BASELINE_SECTION}
We develop a baseline model that can capture the behavior of the devices efficiently with low computational complexity using the steady-state probabilities of the events $p\left(S_{t_{ss}}^{(n)}\right)$ as follows 
%\textcolor{red}{MS: I removed the "given St" from the left hand side}
%%%%% correct %%%%%%%
\begin{align} %\label{activation prediction}
   \Pr\left( A_{t+1}^{(k)}=1\right)&=1- \sum_{\mathcal{\textbf{S}}_t} \prod_{n=1}^N (1-q_{nk})^{\mathcal{S}_t^{(n)} } p\left(S_{t_{ss}}^{(n)}\right),  \label{Steady_State_activ}
\end{align}
where
\begin{equation}
   p\left(S_{t_{ss}}^{(n)}\right)=
      \begin{cases}
        \frac{\epsilon_0^{(n)}}{\epsilon_0^{(n)}+\epsilon_1^{(n)}}, & \quad \mathcal{S}_t^{(n)}=0,\\
       \frac{\epsilon_1^{(n)}}{\epsilon_0^{(n)}+\epsilon_1^{(n)}}, & \quad  \mathcal{S}_t^{(n)}=1.
     \end{cases}
     \label{STEADY_EQ}
\end{equation}

The steady-state probabilities of the events, as calculated in~\eqref{STEADY_EQ}, describe how likely each state will be active long enough during the simulation time~\cite{STEADY_BOOK}. We formulate a probability density distribution (PDF) by multiplying the steady-state probabilities the device activation probabilities as in~\eqref{Steady_State_activ}. This PDF describes the probability of a device to be active affected by the steady-state probability of the states. Hence, this distribution gives a simple description of the activation pattern of the devices without performing any forecasting computations. Afterward, the devices are scheduled by the BS according to this distribution. Note that we refer to this scheduling algorithm as the baseline model.

%The steady-state probabilities are multiplied with the device activation probabilities to formulate a distribution that describes how. We formulate a probability density function (PDF) distribution using the probabilities in \eqref{Steady_State_activ}, then 

%%%%%%%%%%%%%%%%%%%%%%%%%%%%%%%%%%%%

\subsection{AoI Compensation}\label{AGE_COMP_SECTION}
We introduce the age parameter $\beta$ to map the priority of scheduling devices that have high AoI. Higher values of $\beta$ mean that the BS gives higher priority to devices that have not transmitted for a long time (i.e, devices with higher AoI). The scheduling priority index for device $k$ at time $t+1$ is thus defined as 
\begin{align} %\label{activation prediction}
   I_{t+1}^{(k)}&=\Pr\left( A_{t+1}^{(k)}=1 \middle| \mathcal{\textbf{S}}_t\right)+\beta~p\left(S_{t_{ss}}\right) \Delta^{(k)}(t) \\
   &=1-\prod_{n=1}^N h(n)+\beta~p\left(S_{t_{ss}}\right) \Delta^{(k)}(t). \label{modified_activation_probabilities}
\end{align}
Instead of sorting the devices according to their probability of activation, the BS sorts the devices according to their index $I$. Then the $L$ devices with the highest index $I$ are scheduled for transmission.

%\textcolor{blue}{Hirley: from here up to the end of this section, there are some repetitions. You dont need to define (31), you can define the optmization problem directly (are there any constraints on $\beta$? should there be?) Then re-organize the discussion around the optmization problem, results, and discussion}
%The BS needs to choose an appropriate value for $\beta$ to control the trade-off between AoI optimality and regret optimality for the devices.

%{
%\color{red}

\begin{figure}[!t]
\centering
\includegraphics[width=1\columnwidth]{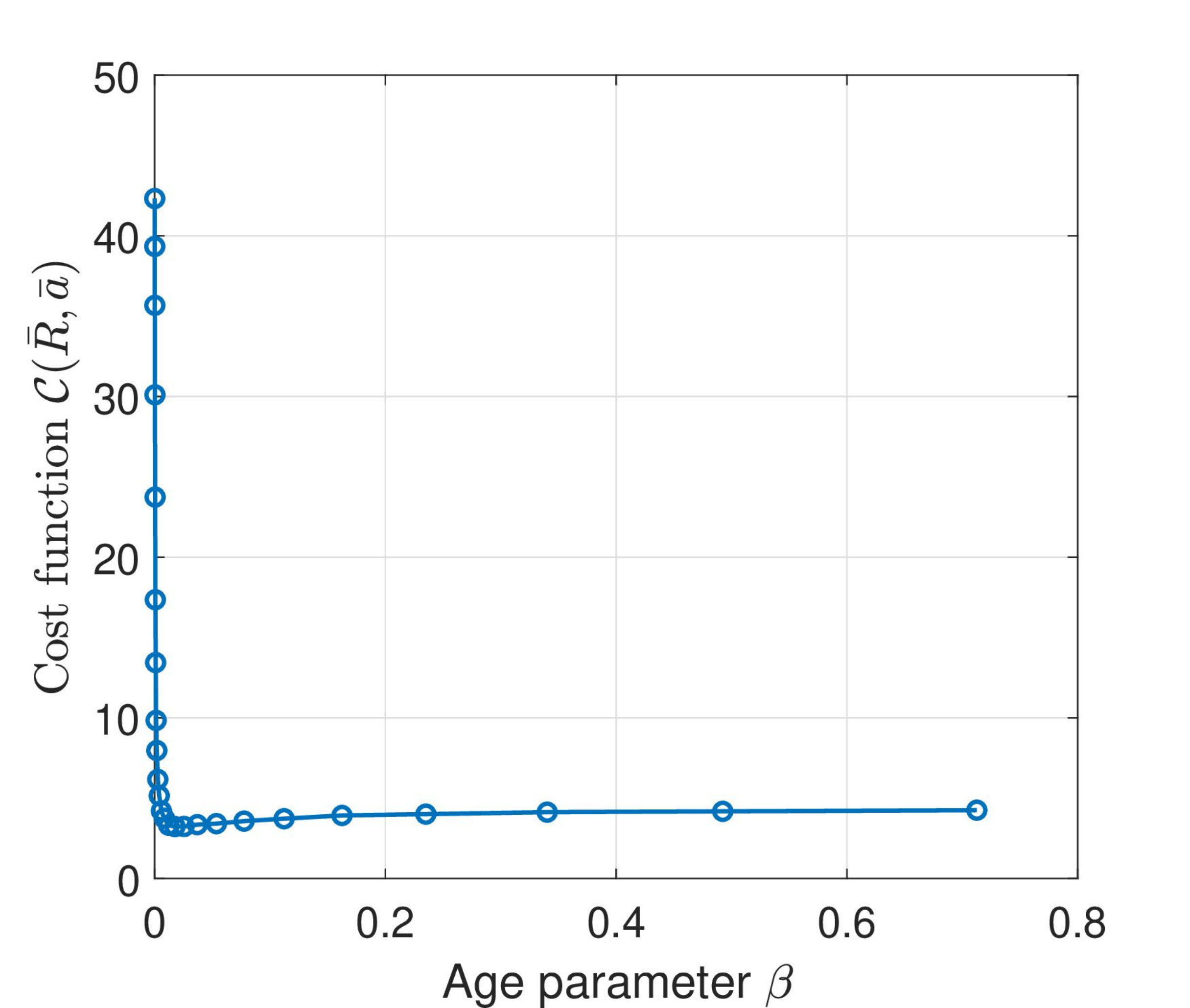}
\centering
%\vspace{2mm}
\caption{Cost function $\mathcal{C}(\Bar{R},\Bar{\Delta})$ for the age parameter $\beta$ optimization.}
%\vspace{-0mm}
\label{Obj_Fun}
\end{figure}

The BS needs to choose an appropriate value for $\beta$ in~\eqref{modified_activation_probabilities} to control the trade-off between the devices' AoI and regret optimalities. This introduces an optimization problem at the BS side, where the cost function $\mathcal{C}(\Bar{R},\Bar{\Delta})$ is defined as the multiplication of the average regret $\Bar{R}$ and the average AoI $\Bar{\Delta}$
\begin{align}
  &\argmin_{\beta} \: \: \mathcal{C} = \Bar{R}\cdot \Bar{\Delta}, \\
  & \: \: s.t. \: \: \: \: \: \:\beta \geq 0.
  \label{Beta_Optim}
\end{align}
As illustrated in Fig.~\ref{Obj_Fun}, we can notice that the cost function is convex and can be optimized easily to get the optimal $\beta$ that lowers down the AoI while maintaining the regret in an appropriate region for a given network setup.

\begin{figure}[!t]
\centering
\includegraphics[width=1\columnwidth]{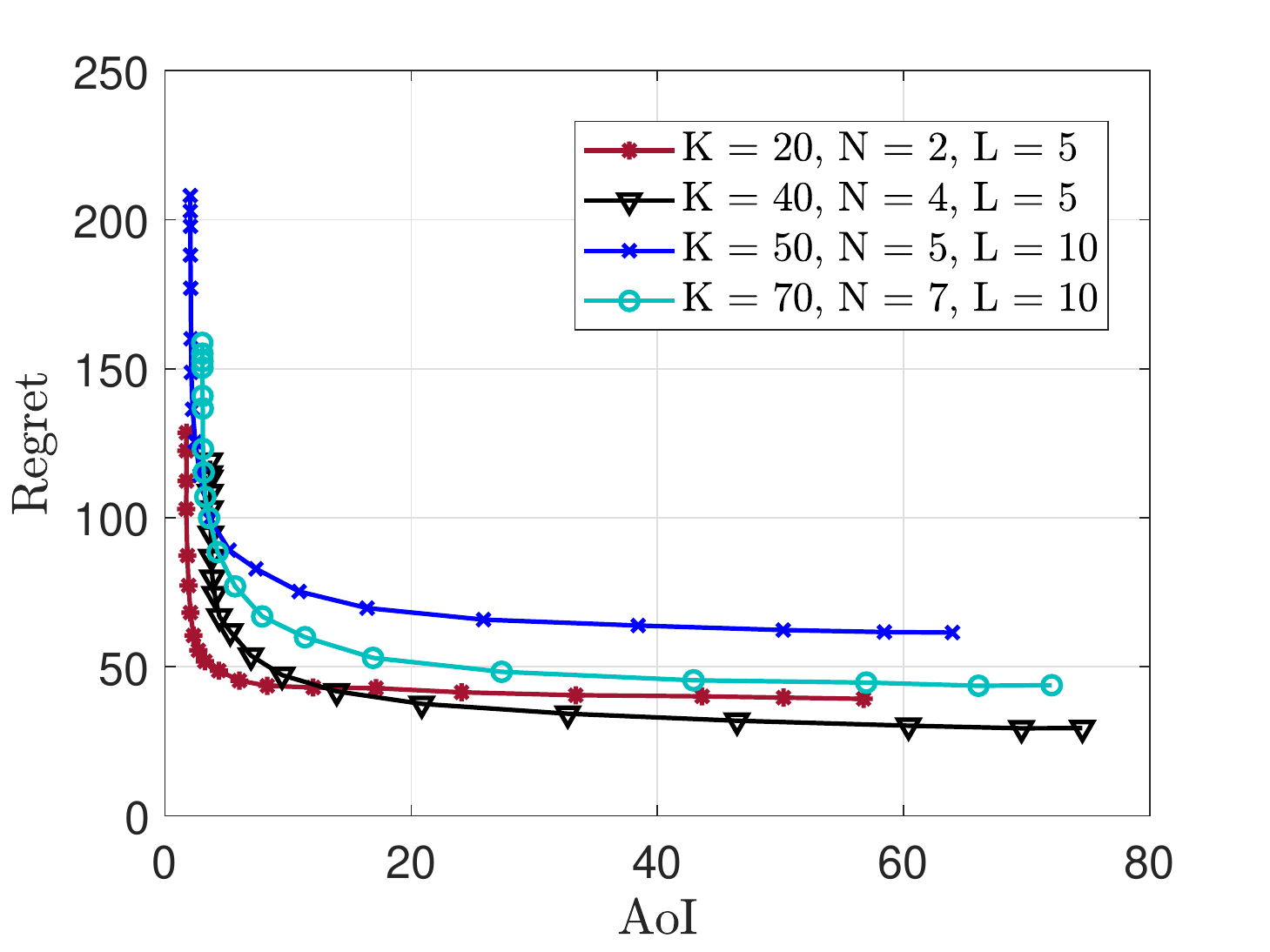}
\centering
%\vspace{2mm}
\caption{Achievable region for the AoI and the regret while applying the AoI compensation for different values of the age parameter $\beta$.}
%\vspace{-0mm}
\label{Ach_Region}
\end{figure}

To address the trade-off between the AoI and the regret, we investigate Fig.~\ref{Ach_Region} that depicts the achievable region for AoI and regret using different values of $\beta$ for different setup of the network (the number of devices, the number of binary events, and the available number of resources). The smaller the network setup $K,N,$ and $L$, the smaller the resulting age and regret. Therefore, each BS needs to optimize its own $\beta$ according to the prior knowledge of the network parameters. If $\beta$ is set to $0$, the scheduling resets to its basic form without the age compensation term (fair regret), whereas if $\beta$ is set to asymptotically $\infty$, the scheduler will act as round-robin, where the resources are distributed equally among the devices (fair age).
%We can notice the trade-off between addressed trade-off between the AoI and the regret. For instance, if we choose extremely small $\beta$, the scheduler will neglect the AoI and just schedule the devices according to the prediction of the forward algorithm. On the other hand, if we choose very large $\beta$, the weight of the regret will be negligible compared to that of the AoI and the scheduler will allocate the resources to the devices that have higher age without taking into consideration the prediction of the forward algorithm. Minimizing the cost function in \eqref{Objective_Function} gives the most beneficial value of $\beta$ that results in low age while keeping the regret and the system usage within an appropriate range

%%%%%%%%%%%%%%%%%%%%%%%%%%%%%%%%%%%%
%}%end-red 

\section{Online Learning Based on Model Estimation}\label{ONLINE_LEARNING_SUBSECTION}
The forward algorithm and the HMM mainly depend on prior knowledge of the hyperparameters of the model, namely, the transition state probabilities for each event and the activation probabilities when affected by active events. Sometimes, it is difficult to have prior knowledge of these parameters. Therefore, the BS aims at estimating the hyperparameters of the model using only the possible observations from the real-time model. Next, we present the estimation algorithm for both $q_{nk}$ and $\epsilon$.
%

%%%%%%%%%%%%%%%%%%%%%%%%%%%%%%%%%%%%

%\vspace{2mm}
The activation probabilities of device $k$ at time instant $t$ given the set of states $\textbf{S}_t$ are the set of values that result in an activation pattern that is as close as possible to the actually observed activation pattern $A_{t}^{(k)}$.
To estimate $q_{nk}$, we formulate the following likelihood maximization formula
\begin{equation}\label{arg_max_q}
\textbf{q}_{nk}^*=\argmax_{\textbf{q}_{nk}}~\prod_{t=1}^T \Pr\left( A_{t}^{(k)}=b_k \middle| \textbf{S}_t^* \right), 
\end{equation}
where $b_k \in \left\lbrace 1,0\right\rbrace$, $\Pr\left( A_{t}^{(k)}=b_k \middle| \textbf{S}_t^* \right)$ is calculated as follows
\begin{equation}
   \Pr\left( A_{t}^{(k)}=b_k \middle| \textbf{S}_t^* \right)=
      \begin{cases}
        1-\prod_{n=1}^N (1-q_{nk})^{\mathcal{S}_t^{^*(n)}},& \: b_k=1,\\
       \prod_{n=1}^N (1-q_{nk})^{\mathcal{S}_t^{^*(n)} },& \: b_k=0,
     \end{cases}
\end{equation}
with the constraint $0<q_{nk}<1$. Note that~\eqref{arg_max_q} can be solved via geometric programming which can be solved for each device $k$ using any programming tool, such as fmincon, which is available in Matlab, or cvx (available in both Matlab and Python)~\cite{cvx}, or even using a basic exhaustive search algorithm to find the solution of the optimization problem. In this context, the cvx tool is considered the best fit for such complex problems with multiple local maxima, where it can solve geometric programming problems efficiently. However, the optimization problem relies on predicting the most likely hidden state $\mathcal{S}_t^*$ from~\eqref{eq:most_likely_S} using the forward algorithm, which uses the actual hyperparameter values $q_{nk}$ and $\epsilon$. This problem can be solved iteratively using the Baum-Welsh algorithm~\cite{HMM2}.
%
%We use the Baum-Welsh algorithm (an expectation maximization method) to estimate $\epsilon$ \cite{HMM2}. 

The Baum-Welsh method relies on the forward-backward algorithms, where at time instant $t$, it estimates the expected number of visits of each state and the number of transitions from state $S_i$ to state $S_j$ during the time period T ($0 \leq T \leq t$). Afterward, it exploits the number of visits and transitions to generate an estimate of $\epsilon^*$. The estimated temporal transition probabilities $\epsilon^*$ along with the previous estimate of $q_{nk}^*$ are used to predict the most likely hidden state, which will be used to update the estimate of $q_{nk}^*$. These iterations are repeated until convergence (desired error threshold). It is expected that the Baum-Welsh algorithm\footnote{A more interested reader can refer to~\cite{HMM2} for more details about the Baum-Welsh expectation-maximization algorithm.} converges after a limited number of iterations $Z$ according to the complexity of the model. After convergence, we can exploit the estimated hyperparameter values $q_{nk}^*$ and $\epsilon^*$ to perform resource allocation for the devices. After initializing $q_{nk}(0)$, $\epsilon_0^{(n)}(0)$ and $\epsilon_1^{(n)}(0)$, we apply the following equations that illustrate the expectation-maximization estimation procedure
\begin{align}
    %&\Pr\left( \mathcal{S}_{t+1}^{(n)}=0 \middle| \mathcal{S}_t^{(n)}=1\right) =\epsilon_0^{(n)}(i-1), \\
    %&\Pr\left( \mathcal{S}_{t+1}^{(n)}=1 \middle| \mathcal{S}_t^{(n)}=0\right) =\epsilon_1^{(n)}(i-1), \\
    %&\Pr\left( \mathcal{S}_{t+1}^{(n)}=0 \middle| \mathcal{S}_t^{(n)}=0\right) =1-\epsilon_1^{(n)}(i-1),  \\
    %&\Pr\left( \mathcal{S}_{t+1}^{(n)}=1 \middle| \mathcal{S}_t^{(n)}=1\right) =1-\epsilon_0^{(n)}(i-1),\\
    &\textbf{S}_t^*(i)=\argmax_{\textbf{S}_t} ~ p(\mathcal{\textbf{S}}_t,\textbf{A}_{1:t})\bigg\rvert_{q_{nk}=q_{nk}^*(i-1),\epsilon_{b_k}^{(n)}=\epsilon_{b_k}^{^*(n)}(i-1)},\\%\label{eq:most_likely_S_i}
    &\Pr\left( A_{t}^{(k)}=b_k \middle| \textbf{S}_t^*(i) \right)= \nonumber\\
      &\qquad\begin{cases}
        1-\prod_{n=1}^N (1-q_{nk})^{\mathcal{S}_t^{^*(n)}(i)},& \: b_k=1,\\
       \prod_{n=1}^N (1-q_{nk})^{\mathcal{S}_t^{^*(n)}(i)},& \: b_k=0,
     \end{cases}\\
    &\textbf{q}_{nk}^*(i)=\argmax_{\textbf{q}_{nk}}~\prod_{t=1}^T \Pr\left( A_{t}^{(k)}=b_k \middle| \textbf{S}_t^*(i) \right).
    \label{ITER_OPTIMI}
\end{align}
%\vspace{-3mm}
%\begin{equation}\label{eq:most_likely_S_i}
%\textbf{S}_t^*(i)=\argmax_{\textbf{S}_t} ~ p(\mathcal{\textbf{S}}_t,\textbf{A}_{1:t}),
%\end{equation}
%\begin{align}   
%\Pr\left( A_{t}^{(k)}=b_k \middle| \textbf{S}_t^*(i) \right)&= \nonumber\\
%&      \begin{cases}
%        1-\prod_{n=1}^N (1-q_{nk})^{\mathcal{S}_t^{^*(n)}(i)},& \: b_k=1,\\
%       \prod_{n=1}^N (1-q_{nk})^{\mathcal{S}_t^{^*(n)}(i)},& \: b_k=0,
%     \end{cases}
%\end{align}
%\begin{equation}\label{arg_max_q_i}
%\textbf{q}_{nk}^*(i)=\argmax_{\textbf{q}_{nk}}~\prod_{t=1}^T \Pr\left( A_{t}^{(k)}=b_k \middle| \textbf{S}_t^*(i) \right), 
%\end{equation}

In fact, this learning process requires enough number of observations to ensure an accurate estimation procedure. If the BS has prior knowledge to a number of observations that is large enough to perform the estimation, we refer to it as FU-offline learning. On the other hand, applying this iterative expectation-maximization procedure at each time-step converts the ordinary algorithm to an online version of the FU algorithm. First, the BS collects the observations at time instant $t$, where it utilizes them to iteratively estimate the model hyperparameters $\textbf{q}_{nk}$ and $\epsilon$. Afterward, it predicts the activation pattern probability of each device at time instant $t+1$ using the forward algorithm. Moreover, it optimizes the age parameter $\beta$ to compensate for the age of the devices that experience high age. Finally, the BS allocates the resources to the devices with the highest priority index. We refer to this procedure as online learning-enhanced AoI, which is depicted in Algorithm~\ref{alg1}.

%as
%\begin{equation}\label{BW_algo}
%\Bar{\epsilon} = \frac{expected \ number \ of \ transitions \ from \ S_i \ to \ S_j}{expected \ number \ of \ transitions \ from \ S_i}.
%\end{equation}
%%%%%%%%%%%%%%%%%%%%%%%%%%%%%%%%%%%%

\begin{algorithm}[!t]
\SetAlgoLined
$t = 1$.

Define $K$, $N$, $L$, and $Z$.

Initialize the age vectors $\Delta^{(k)}$.

Initialize the regret vectors $R^{(k)}$.

\While{True}{
    Initialize $q_{nk}(0)$, $\epsilon_0^{(n)}(0)$ and $\epsilon_1^{(n)}(0)$.
    
    Collect the observations $A_t$.
    
    \For{i = 1,...,$Z$}{
        $\textbf{S}_t^*(i)=\argmax_{\textbf{S}_t} ~ p(\mathcal{\textbf{S}}_t,\textbf{A}_{1:t})\bigg\rvert_{q_{nk}=q_{nk}^*(i-1),\epsilon_{b_k}^{(n)}=\epsilon_{b_k}^{^*(n)}(i-1)}$.
        
        Update $\epsilon_{b_k}^{^*(n)}(i)$.
        
        $\textbf{q}_{nk}^*(i)=\argmax_{\textbf{q}_{nk}}~\prod_{t=1}^T \Pr\left( A_{t}^{(k)}=b_k \middle| \textbf{S}_t^*(i) \right)$.
        }

    Optimize the age parameter $\beta$.
    
    Compensate $\Pr\left( A_{t+1}^{*(k)}=1 \middle| \mathcal{\textbf{S}}_t^*\right)$ using $\beta$. $I_{t+1}^{(k)}=\Pr\left( A_{t+1}^{(k)}=1 \middle| \mathcal{\textbf{S}}_t\right)+\beta~p\left(S_{t_{ss}}\right) \Delta^{(k)}(t)$.
    
    Allocate the $L$ resources.
    
    Update the age vector $\Delta^{(k)}$ for each device.
    
     Update the regret vector $R^{(k)}$ for each device.
    
    t = t+1.
    }
\caption{Traffic prediction based fast uplink grant algorithm.}
\label{alg1}
\end{algorithm}

\section{Results and Discussion}\label{results}
%\textcolor{blue}{Hirley: please harmonize your nomenclature and then revise the text in this section. Is RA, grant-free, baseline all the same? is fast uplink the same as FU? 
%See comment in fig5/7 the naming changes in the text and figures, make sure it is harmonized and defined only once. 
%Might be good to recall what each method does, e.g. 
%- grant-free: baseline model where all nodes do ... defined in section XY
%- TDMA: round robin allocation .... 
%- FU-x: FU as in [X]
%- FU-y: FU with limited information ... defined in Section Z. 
%..... 
%So that the person reading it remembers all the methods and can easily spot the differences
%}

\begin{figure*}[!t]
    \centering
    \subfloat[Regret]{\includegraphics[width=0.5\textwidth,trim={0.1 0 0 0},clip]{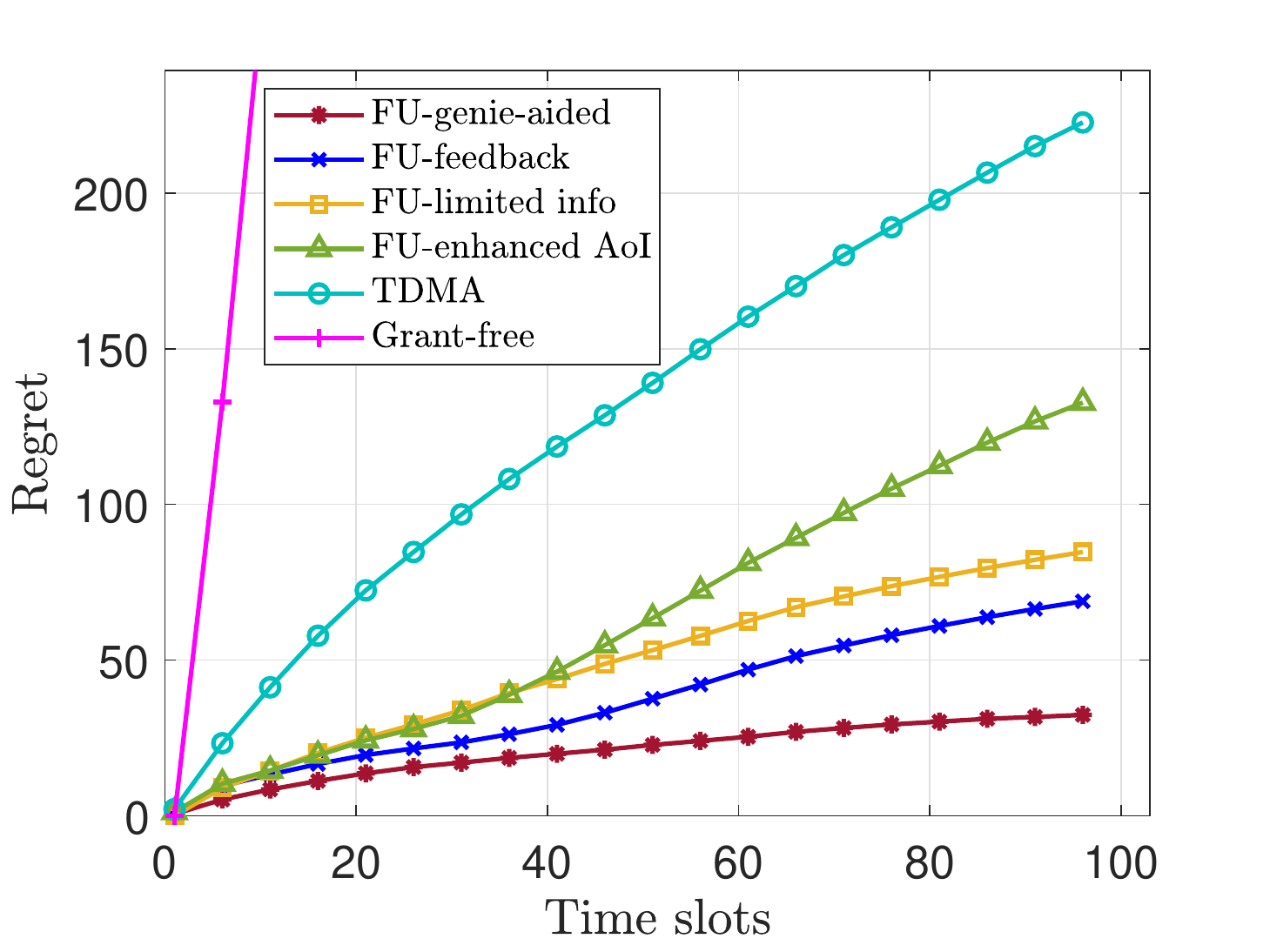}}
    \hskip -2.28ex
    \subfloat[AoI]{\includegraphics[width=0.5\textwidth,trim={0.1 0 0 0},clip]{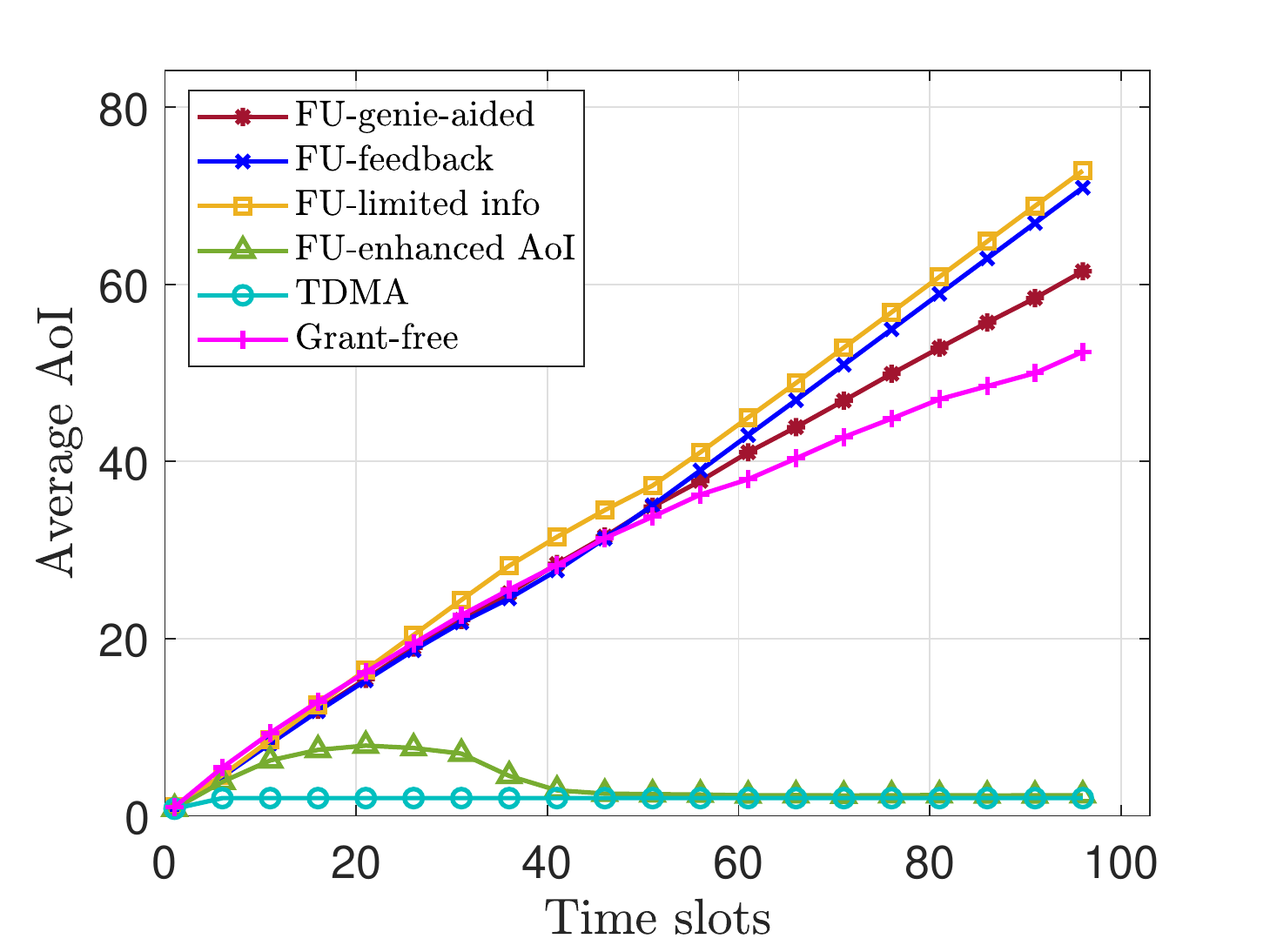}}
    %\hskip -2.28ex
    \caption{Regret and AoI evaluation. $K=50$ sensors, $N=5$ events, and $L=10$ available frequency resources.}
    \label{INIT_FIG} %\vspace{-0mm}
\end{figure*}

In this section, we present the simulation results of the proposed FU algorithm based on the forward algorithm and the further discussed extensions. We consider a setup of a single BS with $L=10$ available frequency resources at each time instant and $K=50$ sensors affected by $N=5$ Markovian events. The temporal state transition probabilities are $\epsilon_0^{(n)}$ and $\epsilon_1^{(n)}$ are uniformly distributed on the interval $[0,0.5]$. Note that, low values of $\epsilon$ result in forcing the events to be active for longer times and cause congested traffic. Meanwhile the activation probabilities $q_{nk}\in[0,1]$. We present a detailed comparison between the proposed algorithms and some of the existing models. For instance, we discuss the GF, where the active devices send a request to the BS using a random preamble, and the TDMA, where round-robin is followed to schedule the resources for the devices. In addition, we present the FU-genie-aided that refers to the case in which the states of the events are assumed to be perfectly known to the BS. Herein, the FU-limited info refers to the scenario in which the BS observes only the activation of the scheduled sensors. Meanwhile, in the FU-feedback, the BS is allowed to also observe the activation of the devices that were not scheduled through a feedback signal.
The FU-baseline is presented as the low computational version of the FU algorithm as presented in~\ref{BASELINE_SECTION}. The term FU-enhanced AoI corresponds to the FU algorithm after performing the age compensation as discussed in~\ref{AGE_COMP_SECTION}.
Finally, FU-offline learning corresponds to applying the estimation algorithm discussed in~\ref{ONLINE_LEARNING_SUBSECTION} while assuming a prior knowledge of enough observations offline to be used to estimate the model hyperparameters, whereas online learning-enhanced AoI is the online version of the presented algorithm, where no prior information is assumed to be known and age compensation is applied as discussed in algorithm~\ref{alg1}. Table~\ref{TABLE_III} illustrates the parameters used in the simulation.

\begin{table}[t!]
\centering
\caption{The parameters used in the simulation setup.}
\label{TABLE_III}
\begin{tabular}{@{}cc@{}|cc@{}cc@{}}
\toprule
\textbf{Parameter} & \textbf{Value}  &  \textbf{Parameter} & \textbf{Value} \\ \midrule
$K$ & $50$  &  $N$ & $5$ \\
$L$ & $10$  &  $T$ & $100$ \\
$Z$ & $40$  &  $\beta$ & $0.0233$ \\
$\epsilon_0^{n}$,$\epsilon_1^{n}$ & $[0,0.5]$  &  $q_{nk}$ & $[0,1]$ \\
\bottomrule
\end{tabular}
\end{table}

%\vspace{0.5mm}

%%%%%%%%%%%%%%%%%%%%%%%%%%%%%%%%%%%%

% \subsection{Simulation Results}

Fig.~\ref{INIT_FIG} demonstrates the regret and the average AoI performance metrics when applying the discussed schedulers. In Fig.~\ref{INIT_FIG}-(a), we evaluate the regret function, where the FU-feedback scheme significantly outperforms both GF and TDMA. Specifically, when applying the proposed FU-feedback scheme, the regret function is reduced to 4 times less than the regret in the case of TDMA and 50 times less than the regret of GF due to the high number of collisions in GF. Moreover, the FU-limited info scheme has close results in terms of the regret to the genie-aided model which assumes perfect knowledge of the events. The feedback version of the FU algorithm exploits the cost of having imperfect information about the activation of the devices, which reflects on the resulting regret. However, the performance is still close to that of the genie-aided model and outperforms existing models (GF and TDMA).

Fig.~\ref{INIT_FIG}-(b) shows the average AoI per device, where the proposed FU-feedback scheme has relatively higher ages when compared to GF and TDMA, which motivates the need for an enhanced AoI version of the FU algorithm. In addition, we calculate the system usage using~\eqref{USAGE_EQ}, where the FU-feedback achieves nearly a $0.95$ system usage, which indicates that the BS has successfully allocated $95\%$ of the resource to the transmitting devices. Hence, the proposed scheme is more efficient than TDMA which uses only $78\%$ of the resources, and the GF that has only $50\%$ of system usage due to the high number of collisions.

%\begin{table}[!t]
%\centering
%    \caption{Performance evaluation for 5 events observed by 100 sensors competing for 10 transmission slots.}
%	\label{Perform}
%\begin{tabular}{|l|l|l|l|}
%\hline
%Algorithm & Regret & Usage & AoI\\ 
%\hline
%Genie-aided & 32 & 0.97 & 64.2\\ 
%\hline
%FU-FB & 57 & 0.95 & 73.6\\
%\hline
%Limited-Info & 78 & 0.93 & 76.6\\ 
%\hline
%Enhanced-AoI & 110 & 0.91 & 02.3\\ 
%\hline
%Estimated-FU & 69 & 0.92 & 73.1\\ 
%\hline
%TDMA & 230 & 0.78 & 02.0\\ 
%\hline
%RA & 2737 & 0.00 & 54.5\\ 
%\hline                              
%\end{tabular} %\vspace{-0mm}
%\end{table}

%%%%%%%%%%%%%%%%%%%%%%%%%%%%%%%%%%%%

Solving the optimization problem in~\eqref{Beta_Optim} renders $\beta = 0.0233$ as the optimal value for the addressed setup. The BS applies the age parameter $\beta$ to address the fairness issue. Fig.~\ref{INIT_FIG} shows the age enhancement which results from applying the fairness parameter $\beta=0.0233$ while scheduling the devices. The average age per device for the FU-enhanced AoI is significantly improved when compared to the basic implementation with $\beta=0$. The average age per device is much lower than GF and asymptotically almost converges to TDMA as time passes instead of being much higher than TDMA in the case of $\beta=0$. Meanwhile, the FU-enhanced AoI still maintains a significant performance advantage regarding regret and system usage when compared to GF and TDMA.

\begin{figure}[!t]
\centering
\includegraphics[width=1\columnwidth]{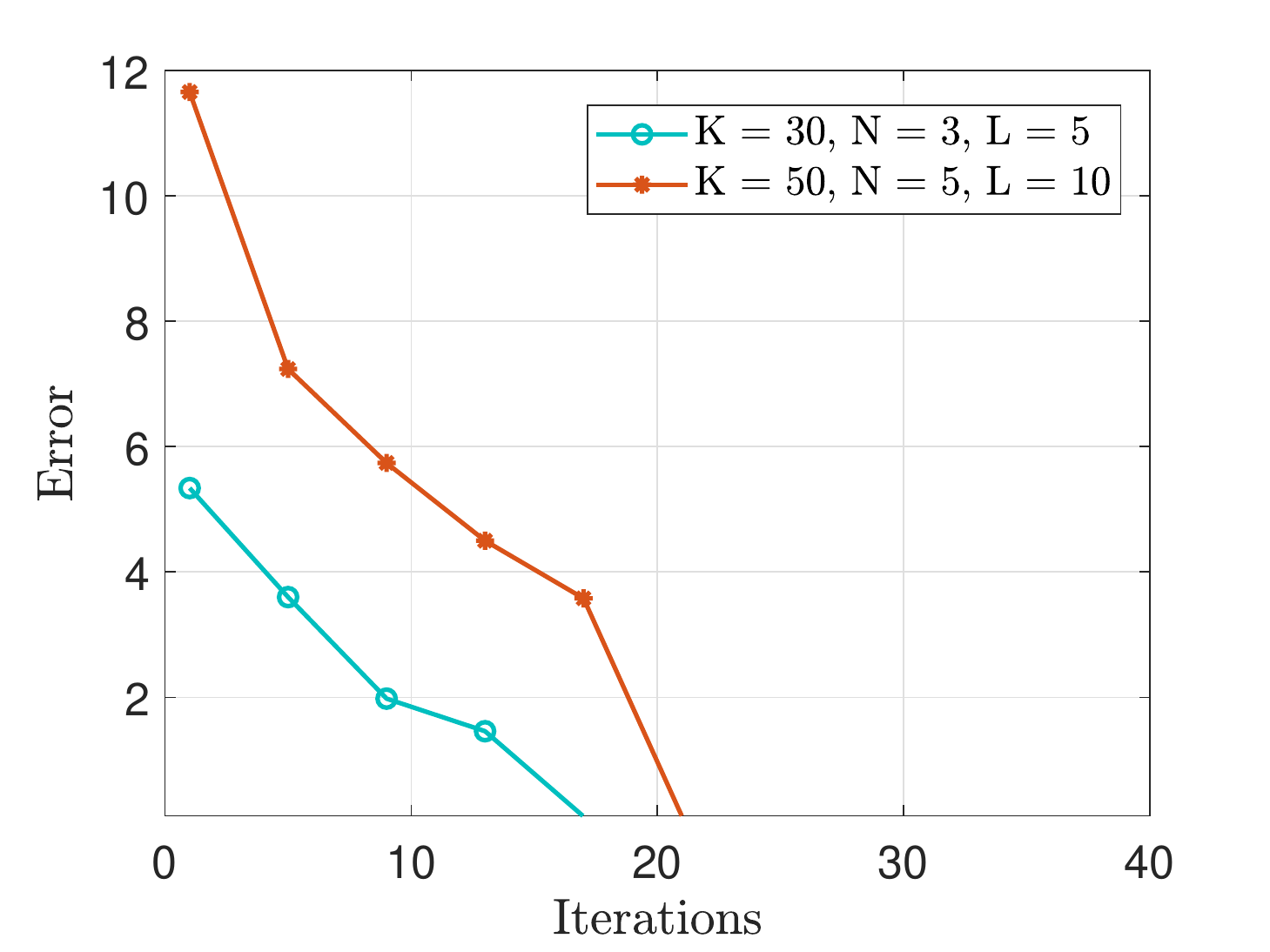}
\centering
%\vspace{2mm}
\caption{The convergence of estimation of the model hyperparameters using the Baum-welsh algorithm.}
%\vspace{-0mm}
\label{Convergence}
\end{figure}

\begin{figure}[]
    \centering
    \subfloat[Regret]{\includegraphics[width=0.5\textwidth,trim={0.1 0 0 0},clip]{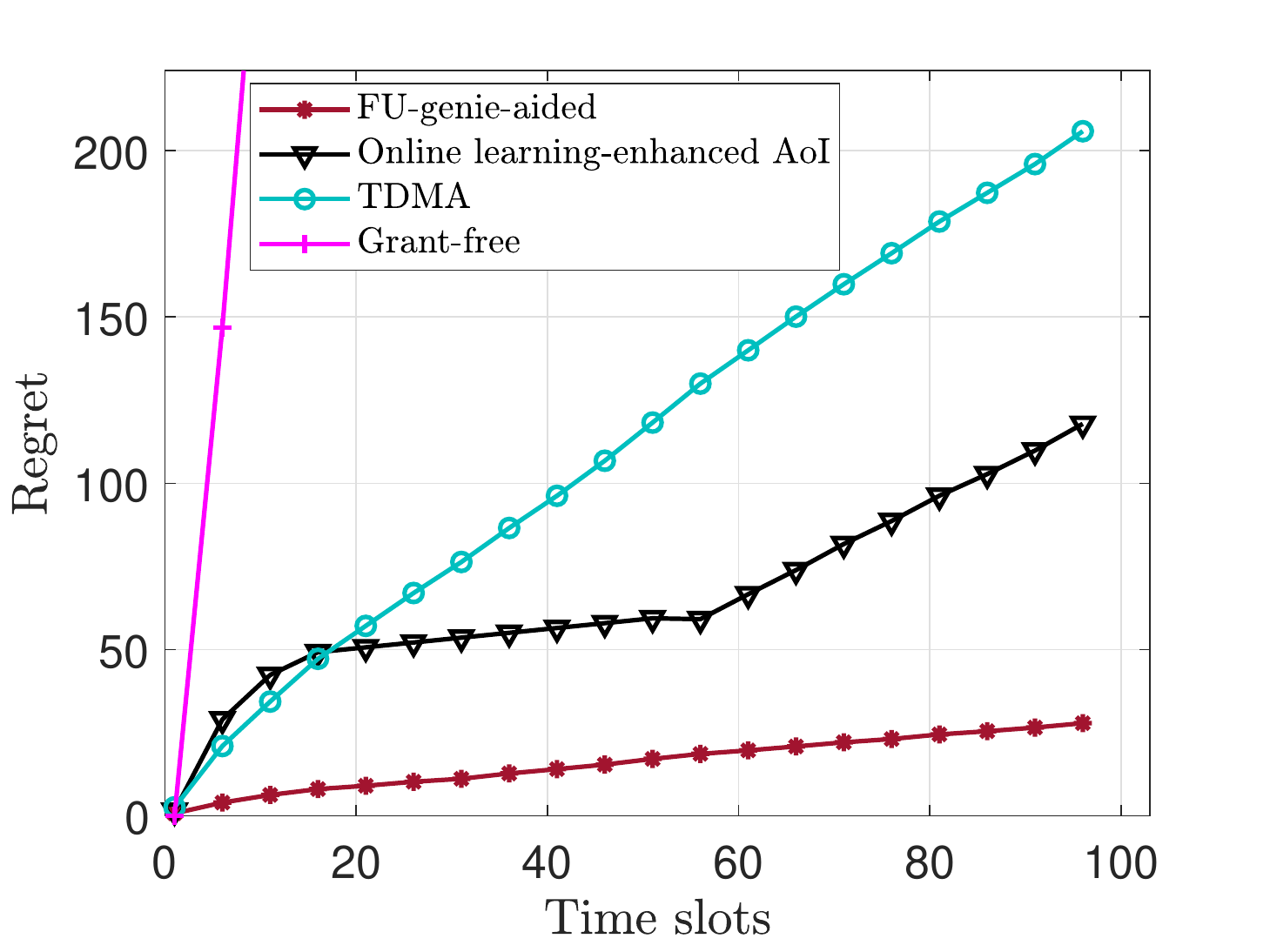}}
    \hskip -2.28ex
    \subfloat[AoI]{\includegraphics[width=0.5\textwidth,trim={0.1 0 0 0},clip]{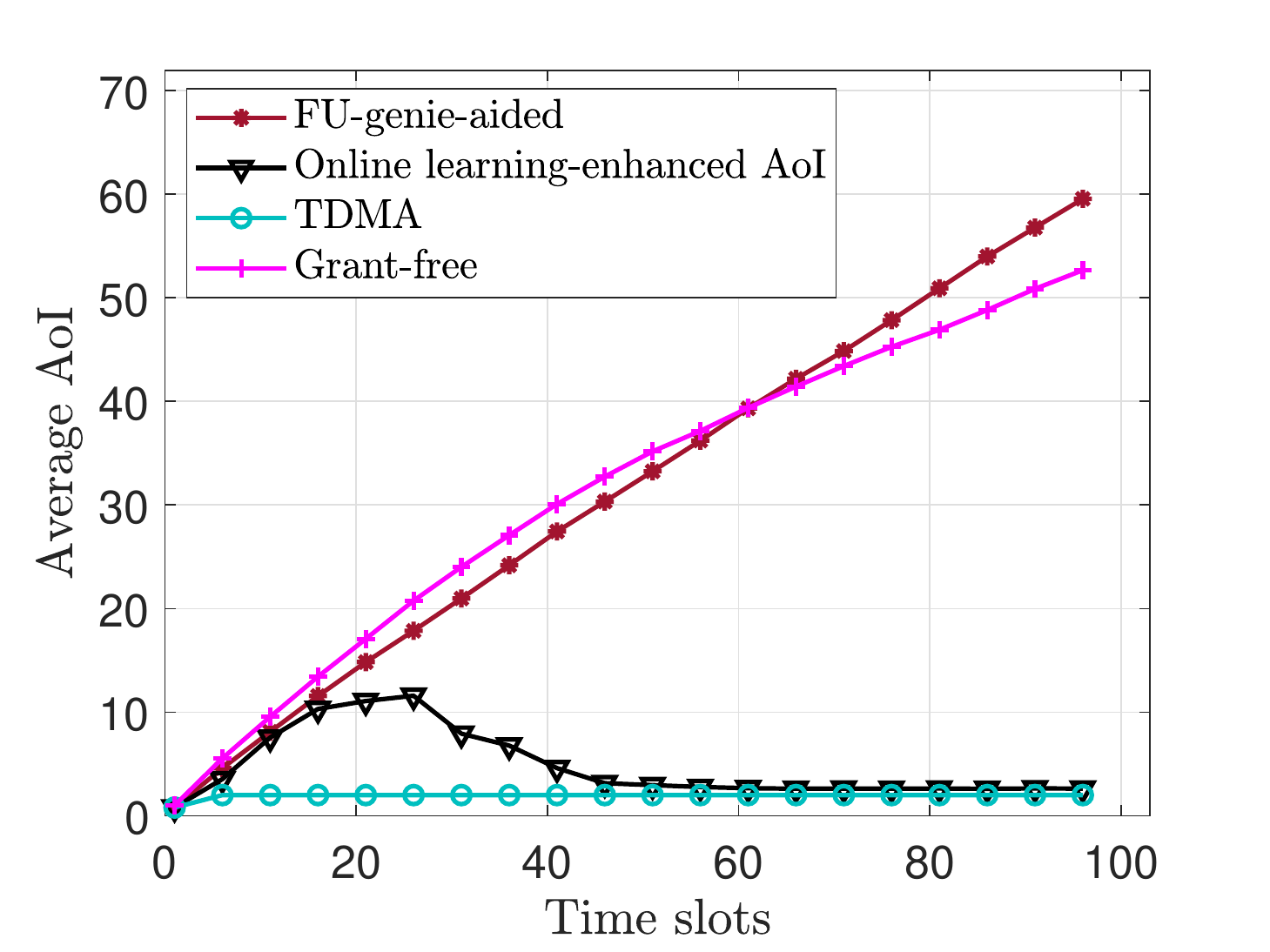}}
    %\hskip -2.28ex
    \caption{Regret and AoI evaluation of the online learning-enhanced AoI algorithm. $K=50$ sensors, $N=5$ events, and $L=10$ available frequency resources.}
    \label{ONLINE_LEARNING_FIG} %\vspace{-0mm}
\end{figure}

Fig.~\ref{Convergence}
illustrates the convergence of the estimated hyperparameter values $q_{nk}^*$ and $\epsilon_{b_k}^{^*(n)}$. The error is measured as the difference between the true regret of the forward algorithm using the true hyperparameter values $q_{nk}$ and $\epsilon$ and the regret resulting from scheduling the resources for the devices using the estimated hyperparameter values. We initialize the values of $q_{nk}^*$ and $\epsilon_{b_k}^{^*(n)}$ and run the iterative optimization algorithm as described in section~\ref{ONLINE_LEARNING_SUBSECTION}. We solve~\eqref{arg_max_q} for each device using both exhaustive search and CVX, where exhaustive search results in a more accurate estimation, while CVX is much simpler and more efficient in terms of estimation time. Afterward, we run the Baum-Welsh algorithm for 40 iterations, where it convergences to reasonable values for $\epsilon_{b_k}^{^*(n)}$ and $q_{nk}^*$ that truly describe the observations.
We can notice the convergence of the model hyperparameters after looping the algorithm for a sufficient number of iterations. Typically, the convergence is significantly faster for a small setup of the system model as the number of states and devices controls the number of the hyperparameters to be estimated.
We run the mentioned estimation procedure to be used in the learning algorithm offline (FU-offline learning) and online (online learning-enhanced AoI), where the former assumes prior knowledge of enough number of observations to run the estimation upon it, whereas the latter runs the estimation algorithm online while accumulating the observations.

Fig.~\ref{ONLINE_LEARNING_FIG} shows the performance evaluation of the online learning-enhanced AoI algorithm in terms of regret and average AoI, respectively. As the algorithm has no prior knowledge about the states and the hyperparameters of the model, it applies the forward algorithm and the age compensation strategy based on the given set of previous observations collected at each time step. We can see in Fig.~\ref{ONLINE_LEARNING_FIG}-(a) that the behavior of the algorithm is not efficient in the initial time steps as there are not enough observations that can describe the model and correctly estimate the model hyperparameters. Afterward, the hyperparameters estimation gets better (almost after 16 time instants) as the model collects a suitable amount of observations that truly describe the model and are used efficiently in the estimation procedure. In Fig.~\ref{ONLINE_LEARNING_FIG}-(b), the algorithm experiences a large AoI compared to the TDMA in the initial time steps, where the age compensation strategy optimizes the age parameter $\beta$ assuming that the prediction results are efficient enough to compensate the true high age devices. %Afterward, the online-learning algorithm collects enough observation to perform an efficient prediction, where the age compensation strategy almost captures the TDMA AoI after 40 time instants.
Afterward, the online learning-enhanced AoI algorithm collects enough observation to efficiently predict the model hyperparameters, where the age compensation strategy almost captures the AoI of the TDMA after 40-time instants.

\begin{figure}[t!]
    \centering
    \subfloat[Regret]{\includegraphics[width=0.44\textwidth,trim={0.1 0 0 0},clip]{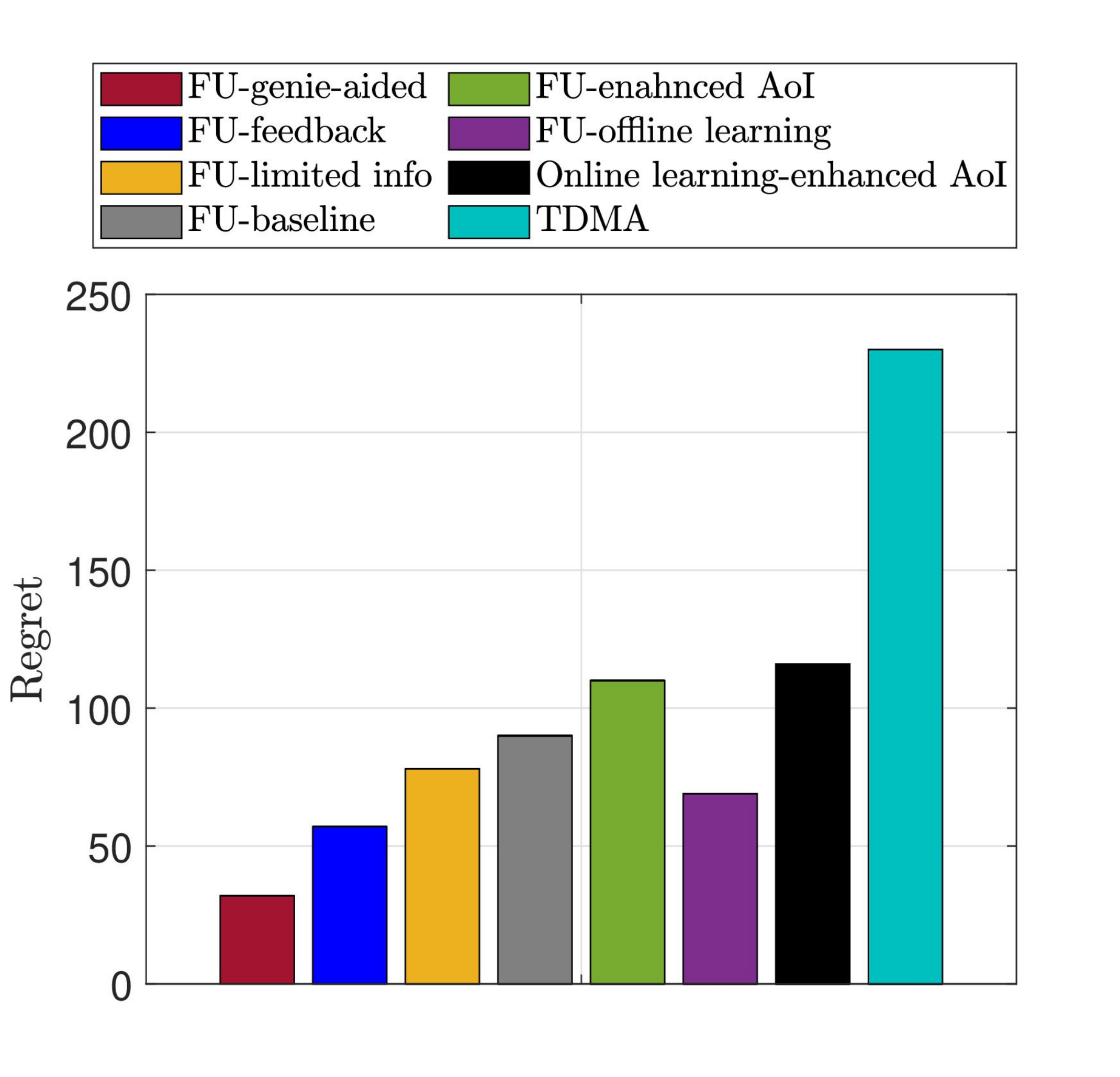}}
    \hskip -2.28ex
    \subfloat[AoI]{\includegraphics[width=0.44\textwidth,trim={0.1 0 0 0},clip]{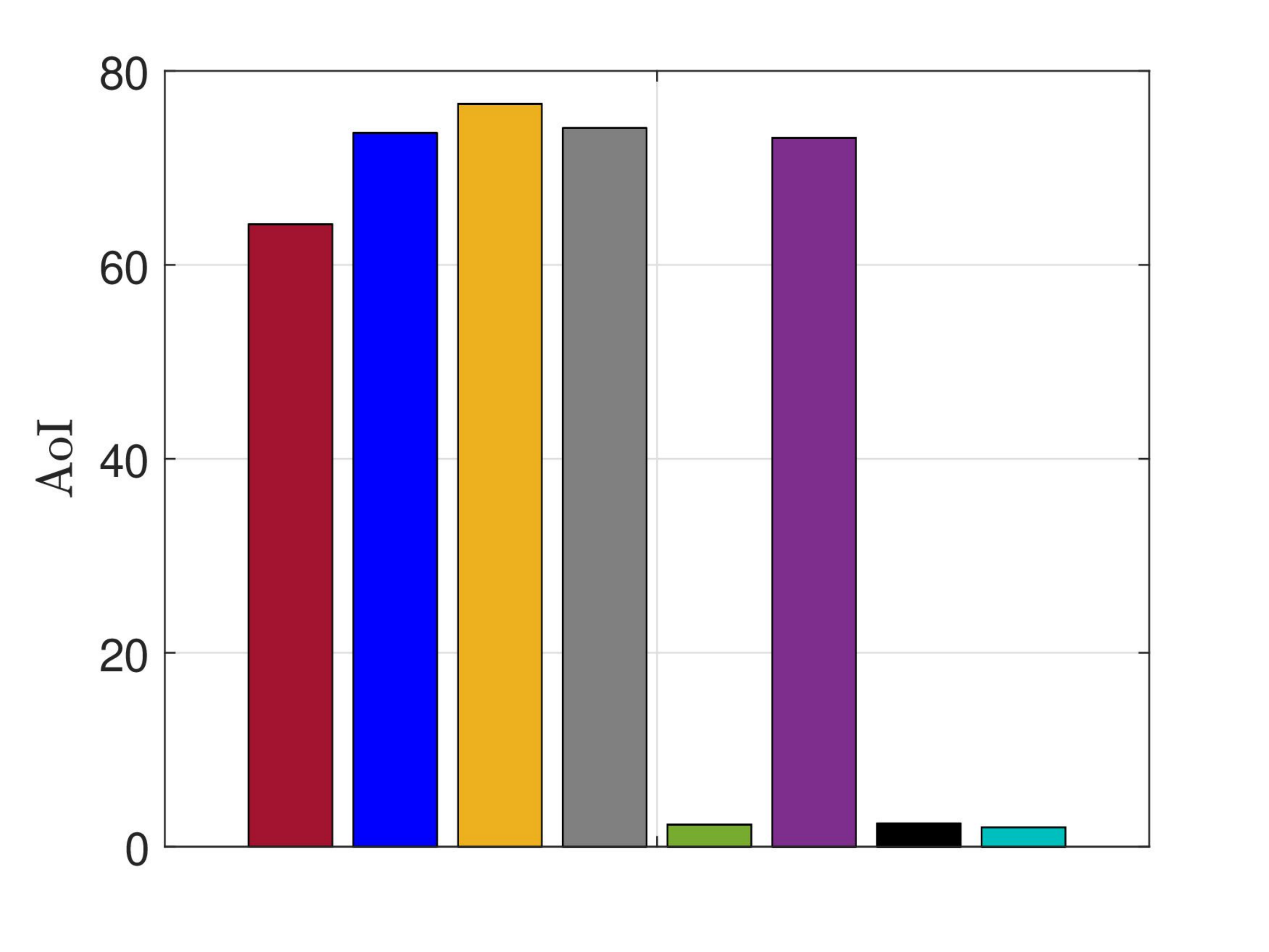}}
    \hskip -2.28ex
    \subfloat[System usage]{\includegraphics[width=0.44\textwidth,trim={0.1 0 0 0},clip]{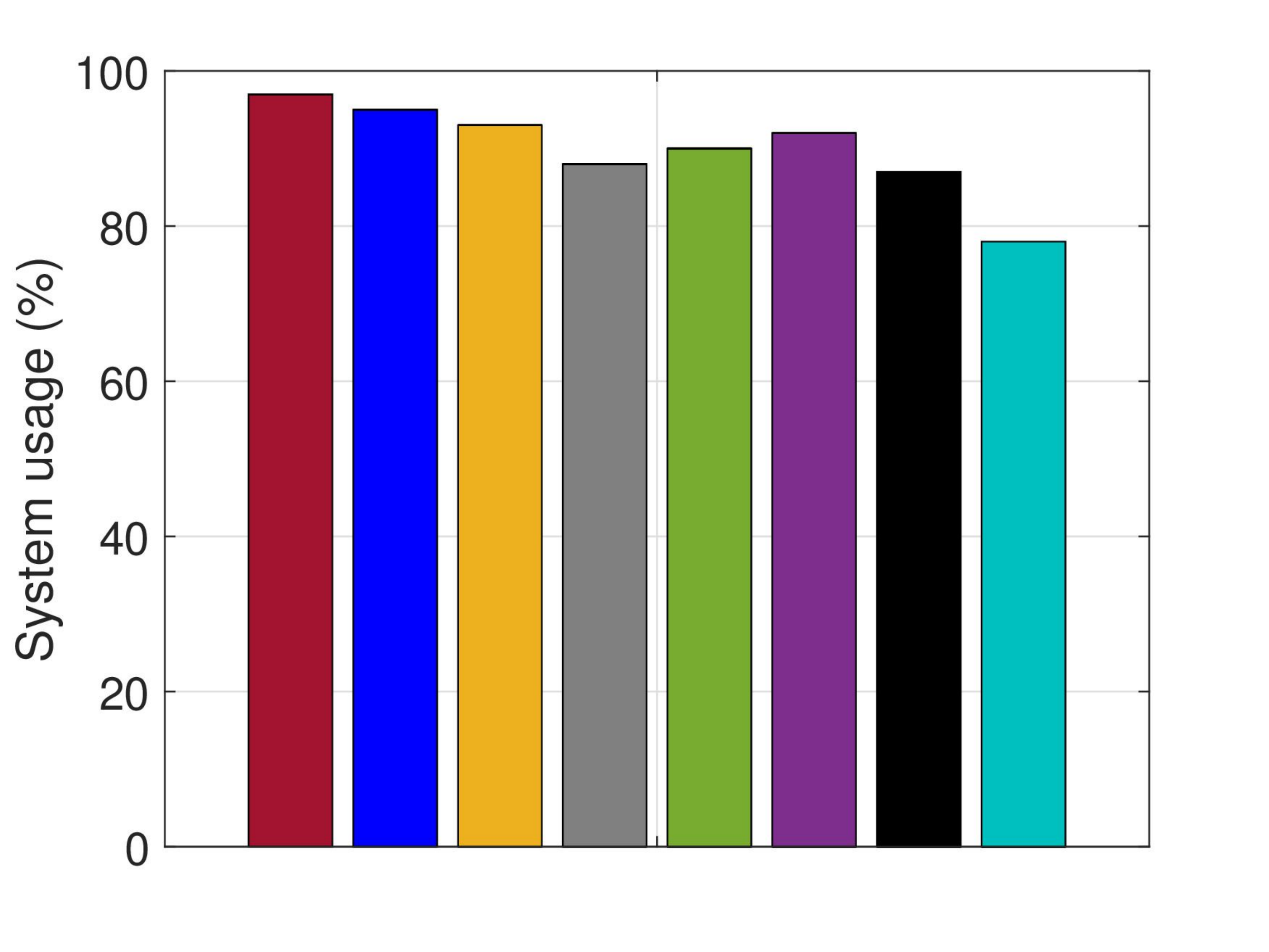}}
    %\hskip -2.28ex
    \caption{Regret, AoI and system usage evaluation. $K=50$ sensors, $N=5$ events, and $L=10$ available frequency resources.}
    \label{BAR_FIG} %\vspace{-0mm}
\end{figure}

Fig.~\ref{BAR_FIG} summarizes the regret, AoI, and system usage performance metrics when applying the proposed resource allocation schemes. It is worth mentioning that the GF results are omitted from the bar plots as it has extremely poor performance compared to all other schemes due to high number of collisions, and this would affect the comprehensive comparison of the schemes on the plots (namely, on the regret bar plot). It results in regret of around $3000$, an AoI of $52$, and system usage of $65\%$. The FU-feedback achieves a reduced regret to $50$ times less than the GF and a slightly less system usage than the FU-genie-aided case with $2\%$ difference. The TDMA has the best AoI results as it is considered as the fair age scheduler. Therefore, age compensation is applied within the FU-enhanced AoI algorithm that captures the AoI of the TDMA of $2.3$ at the expense of slightly higher regret, where it has a $40$ more regret than the FU-feedback. However, it still outperforms the regret and the system usage of TDMA and GF schemes. We can observe that the FU-baseline achieves $3$ times lower regret than and $9\%$ higher system usage than TDMA. Therefore, the FU-baseline still outperforms the TDMA and the GF resource allocation schemes regarding regret and system usage with lower computational demands.

%Moreover, we fit the estimated parameters to the scheduling algorithm to calculate the regret, system usage and average AoI of the model.
Moreover, we fit the estimated parameters to the scheduling algorithm to calculate the model's regret, system usage, and average AoI.
We can observe that both FU-offline learning and online learning-enhanced AoI outperform the regret of the TDMA and almost captures the regret of the FU-feedback. In addition, the online learning-enhanced AoI has almost double the regret of the FU-offline learning ($65$ and $120$ for the FU-offline learning and the online learning-enhanced AoI, respectively) as the online version suffers from inaccurate estimation at the beginning of the simulation as there are not enough observations to be used in the estimation, whereas the offline version assumes prior knowledge of enough observations for the estimation. In addition, the online learning-enhanced AoI performs an AoI compensation step after estimating the model hyperparameters, which enables the algorithm to achieve the AoI of the TDMA while preserving the regret to still outperform the TDMA. Finally, There is an interesting analogy between the FU-limited info and the FU-offline learning results, where both algorithms suffer from missing information as the former has limited information about the actual activation of the devices and depends only on its prediction, whereas the latter relies on a collection of past observations to estimate the model hyperparameters.

\section{Conclusions}\label{conclusions} %\vspace{1mm}
%
%In this paper we consider Markovian events that stimulate massive deployment of IoT devices. 
%This paper considers Markovian events that stimulate a massive deployment of IoT devices. 

%\textcolor{red}{MS: better use past tense here}
This paper considers Markovian events which serve to model the activity of the massive deployment of IoT devices.
We proposed an FU algorithm that efficiently predicts the activation pattern of the IoT devices based on the forward algorithm and grants the available resources to the devices with the highest likelihood of activation probabilities. We formulated an optimization problem that compromises a small value of the regret to minimize the AoI of the IoT devices and achieve a desirable degree of fairness. In addition, we formulated an expectation-maximization algorithm based on the Baum-Welsh procedure to estimate the system hyperparameters. Finally, we developed an online-learning version of the proposed scheme. Simulation results showed that the proposed algorithm outperforms the existing models, e.g., TDMA and GF, regarding regret, system usage efficiency,  and AoI.

The proposed algorithms were much simpler than machine learning-based predictors regarding the complexity of the computations. Therefore, the proposed algorithms could be used as traffic predictors in critical applications, e.g., predictive UAV positioning~\cite{U3}, road safety, and other applications with low latency communications demands~\cite{8705373}.

%Efficient traffic prediction could be applied to preemptively allocate resources to IoT device via resource allocation schemes such as fast uplink (FU) grant. In this paper, we propose a traffic prediction scheme that efficiently predicts the transmission likelihood of sensors stimulated by Markovian events in massive IoT scenarios. Accordingly resource blocks are allocated to sensors that are most likely to be active at each time slot vian FU grant. The proposed scheme significantly outperforms both RA and TDMA in terms of regret and system usage, and this performance superiority is more obvious in more dense networks. 

%For more dense networks, the average age per device of FU is lower than RA and thus, it also outperforms RA age-wise. However, for less dense networks, the average age per device for the proposed FU scheme is higher than both RA and TDMA. The proposed FU grant scheme renders relatively high average age since sensors with very low transmission likelihood are rarely scheduled. This fairness issue is an open problem for future research where the target could be minimizing the regret function subject to peak age per device constraint. Finally, the benefits of efficient traffic prediction in such scenarios could be extended to the efficient allocation of other types of resources such as in optimum UAV (unmanned aerial vehicle) positioning and network design as envisioned by \cite{U3}.
%
%%\vspace*{-1mm} 

\appendices 

\bibliographystyle{IEEEtran}
\bibliography{IEEEabrv,di}
\end{document}